\documentclass[lettersize,journal]{IEEEtran}
\usepackage{amsmath,amsfonts}
\usepackage{algorithmic}
\usepackage{algorithm}
\usepackage{array}
\usepackage[caption=false,font=small,labelfont=rm,textfont=rm]{subfig}
\usepackage{textcomp}
\usepackage{stfloats}
\usepackage{url}
\usepackage{verbatim}
\usepackage{graphicx}
\usepackage{cite}
\usepackage{amssymb}
\usepackage{cleveref}
\usepackage{booktabs}
\usepackage{overpic}                
\begin{document}

\title{Hybrid Precoding With Low-Resolution PSs for Wideband Terahertz Communication Systems in The Face of Beam Squint}

    \author{Yang Wang,~\IEEEmembership{Student Member,~IEEE,}
            Chuang Yang,~\IEEEmembership{Member,~IEEE,}
            Mugen Peng,~\IEEEmembership{Fellow,~IEEE,}
\thanks{This work was supported in part by the National Key Research and
Development Program of China under Grant 2021YFB2900200, the National Natural Science Foundation of China under Grant 62101059, the Beijing Natural Science Foundation under Grant L223007, the Young Elite Scientists Sponsorship Program by CAST under Grant 2022QNRC001, and the Base Independent Scientific Research Project under Grant NST20230304. (Corresponding author: Chuang Yang.)}
\thanks{The authors are all with the State Key Laboratory of Networking and Switching Technology, Beijing University of Posts and Telecommunications, Beijing 100876, China (e-mail: wygoforever@bupt.edu.cn; chuangyang@bupt.edu.cn; pmg@bupt.edu.cn).}
}

\markboth{Journal of \LaTeX\ Class Files,~Vol.~, No.~, June~2024}%
{Shell \MakeLowercase{\textit{et al.}}: A Sample Article Using IEEEtran.cls for IEEE Journals}


\maketitle
\begin{abstract}
Terahertz (THz) communication is considered one of the most critical technologies for 6G because of its abundant bandwidth. To compensate the high propagation of THz, analog/digital hybrid precoding for THz massive multiple input multiple output (MIMO) is proposed to focus signals and extend communication range. Notably,  considering hardware cost and power consumption, infinite and high-resolution phase shifters (PSs) are difficult to implement in THz massive MIMO and low-resolution PSs are typically adopted in practice. However, low-resolution PSs cause severe performance degradation. Moreover, the beam squint in wideband THz massive MIMO increases the performance degradation because of the frequency independence of the analog PSs. Motivated by the above factors, in this paper, we firstly propose a heuristic algorithm under fully connected (FC) structure, which optimize the digital precoder and the analog precoder alternately. Then we migrate the proposed heuristic algorithm to the partially-connected (PC) architecture. To further improve the performance, we extend our design to dynamic subarrays in which each RF chain is connected to any antenna that does not duplicate. The numerical results demonstrate that our proposed wideband hybrid precoding with low-resolution PSs achieves better performance to the comparisons for both FC structure and PC structure.
\end{abstract}


\section{Introduction}
\IEEEPARstart{W}{ith} the fast development of emerging applications, such as internet of vehicles, virtual reality, digital twins, and holographic imaging, the demands of communication rates will increase from Gbps to Tbps in future 6G. However, with the limited bandwidth, 5G mmWave communication cannot meet the rising requirements\cite{wang2023road,yang20196g,10360222}. The terahertz band (100GHz$\sim$1THz) can provide abundant and continuous bandwidth from tens and up to hundreds of GHz, which can be considered one of the crucial techniques for ultra-high-speed communication in future 6G\cite{10045774,10105223}. The most critical issue restricting the application of THz in 6G is the high free-space loss and atmospheric absorption loss of electromagnetic waves in the THz band, which significantly shortens the propagation distance\cite{akyildiz2022terahertz}. To address this challenge, massive MIMO is employed to achieve a highly directional beam to compensate for the high path loss\cite{9398864}.

With very small wavelength, the THz massive MIMO allows large antennas integrated into the transmitter, i.e., containing 256, 512, or even 1024 antennas at the transmitter\cite{9216613}. Therefore, conventional fully digital precoding that each RF chain is connected only to one antenna is not suitable for THz massive multiple input multiple output (MIMO) due to the high cost and power consumption caused by RF chains\cite{9262080}. To address this hardware challenge, hybrid analog/digital precoding has been proposed as a promising solution for THz massive MIMO, which significantly reduces the number of RF chains while simultaneously supporting multiplexing/multi-user scenarios\cite{wang2023sensing}.

\subsection{Prior Work}
Abstracted by the advantages of hybrid precoding, extensive work has been conducted on the architecture and optimization algorithms of hybrid precoding in the last few years. Some optimization methods, such as those based on orthogonal match pursuit (OMP), manifold optimization, alternating minimization, and singular value decomposition (SVD), have been proposed and applied to hybrid precoding in fully connected (FC) structures or partially connected (PC) structure\cite{el2014spatially, lee2014hybrid, yu2016alternating, li2017hybrid, zhang2018svd}. However, these schemes are based on high/infinite resolution. In wideband THz communication systems, the power consumption of PSs increases sharply with increasing resolution\cite{10438529}. Moreover, the PSs with high/infinite resolution are extremely difficult to manufacture in high-frequency band \cite{10159567}. Obviously, it is impractical to utilize high/infinite resolution PSs in mmWave/THz massive MIMO. Therefore, it is crucial to develop signal processing techniques adapted for low-resolution phase shifters in THz massive MIMO to mitigate performance degradation caused by the low-resolution PSs.

Some researches have already been conducted on hybrid precoding with low-resolution PSs\cite{he2023energy,zhu2023max,nie2023spectrum,sohrabi2016hybrid,chen2017hybrid,lyu2021lattice}. A straightforward method to obtain the analog precoding matrix with low-resolution PSs is directly quantizing the phase of each element in the analog precoding matrix derived from the infinite-resolution hybrid precoding schemes. However, this approach will introduce significant errors, especially for methods based on OMP and SVD because the orthogonality is compromised after quantization. Another effective approach is to iteratively optimize the phase of each PS to progressively approximate the optimal spectral efficiency\cite{sohrabi2016hybrid}. To simplify the iteration process, the optimization objective in \cite{chen2017hybrid, lyu2021lattice} is transformed from the optimization of spectral efficiency into the optimization of the Euclidean distance between the fully digital precoding matrix and the hybrid precoding matrix. However, the schemes in the papers mentioned above are based on the assumption of a narrowband system, without considering MIMO systems with orthogonal frequency division multiplexing modulation (OFDM) which is suitable for wideband.

As for MIMO-OFDM systems, some works have been further researched based on the aforementioned foundation and proposed various hybrid precoding schemes for MIMO-OFDM systems. In \cite{sohrabi2017hybrid}, the authors propose to approach fully digital precoding by optimizing the upper performance bound. However, it is difficult to apply in THz wideband systems due to significant channel frequency selectivity. A hybrid precoding scheme based on tensor decomposition is also proposed in \cite{zilli2021constrained}, which maximizes the sum of effective baseband gains over all subcarriers as well as suppresses the inter-user and intra-user interferences. \cite{li2020dynamic} proposes a dynamic hybrid precoding scheme that dynamically connects each RF chain to a non-overlapping antenna subarray via a switch network and PSs to migrate the performance degradation due to low-resolution PSs. Considering the hardware cost caused by PSs, \cite{yan2022dynamic} and \cite{yu2018hardware} propose new architectures that utilize a small number of fixed PSs instead of a large number of traditional PSs to reduce hardware costs. However, it's almost impossible to achieve the THz massive MIMO with fixed PSs currently. Moreover, these schemes also have limitations in THz wideband systems with strong frequency-selective channels. Because all schemes use an iterative optimization approach by minimizing the Euclidean distance. However, this method is only an approximation under high effective signal-to-noise ratios and results in significant errors under strong frequency-selective channels and low signal-to-noise ratios. Therefore, hybrid precoding with low-resolution PSs under wideband frequency-selective channels remains challenging.

Another challenging task is to compensate for beam squint in THz massive MIMO. Beam squint is an unavoidable problem in THz massive MIMO, which is caused by the frequency-independent characteristics of PSs. There exists many literatures aiming to beam squint issue \cite{9262080,9398864,10333083,zhang2018hybrid,li2020dynamic,9908559}. \cite{9262080} proposes to construct a common channel matrix for wideband hybrid precoding by projecting all subcarriers onto the center frequency. Inspired by the radar beam squint, \cite{9398864} proposes to utilize time-delay lines to enhance the frequency dependence of the analog precoded matrix. \cite{10333083} proposes a iternative minimization method to mitigate the beam squint in a wideband system. Those schemes contribute to the beam squint problem, but do not consider the performance degradation caused by low-resolution PSs. \cite{zhang2018hybrid} leverages the penalty decomposition method to improve the performance of hybrid precoding with low-resolution PSs in mmWave wideband systems. However, the performance is limited under the large bandwidth of THz. Though \cite{li2020dynamic} considers the beam squint caused wideband mmWave systems, the performance of the hybrid precoding scheme can be further improved in THz systems with strong frequency-selective channels and greater bandwidth.

\subsection{Contributions}
Against the questions, this paper directs its focus towards hybrid precoding with low-resolution PSs which considers the impact of beam squint and strong frequency-selective channels. We first transform the optimization objective by partially decoupling the digital precoding and the analog precoding. Based on the transformation, we propose several heuristic algorithms for FC and PC structures by developing an iterative block coordinate descent method. The main contributions of this paper can be summarized as follows.

\begin{enumerate}
	\item[$\bullet$] We first analyze the impact of beam squint in massive MIMO systems with low-resolution PSs. As well known, the low-resolution PSs will cause the beamwidth broadening. The wide beam will ease the beam squint problem. Therefore, whether the beam squint issue is serious under the influence of low-resolution PSs needs further verification. Motivated by this, we utilize the existing algorithm that aims at hybrid precoding with low-resolution PSs to simulate the array gain in different directions in this paper. We observe that the beam squint is still obvious even when the beamwidth is broadened, which suggests that our research on hybrid precoding under the combined influence of broadband and low-resolution PSs is valuable.
        \item[$\bullet$] We propose an alternating optimization algorithm in FC structures for THz wideband systems. Specifically, we first simplify the optimization objective. Unlike previous work, we believe that digital precoding and analog precoding are coupled, and completely decoupling digital and analog precoding into two separate optimization stages makes it difficult to achieve an optimal solution. Therefore, in this paper, we partially decouple analog precoding and digital precoding. Then we design an iterative block coordinate descent method to take turns optimizing the analog precoding and digital precoding. In the update part in analog precoding, we isolate the contribution of each element in the analog precoding matrix to spectral efficiency, which takes the impact of digital precoding into consideration. Based on this, we derive the approximate optimal solution for each PS. In this process, the frequency-selective channel effects are fully considered, which is more suitable for THz wideband systems. 
        \item[$\bullet$] We propose two alternating optimization algorithms in PC structures for THz wideband systems. We consider alternating optimization schemes under two kinds of PC structures\textemdash PC structures with fixed subarrays and PC structures with dynamic subarrays. As for PC structures with fixed subarrays, we port the algorithm proposed in the FC structure to the PC structure with fixed subarrays and give the corresponding initialization scheme. With fewer PSs in the PC structure with fixed subarrays, the complexity of the alternating optimization algorithm is reduced. For PC structures with dynamic subarrays, we propose a row-by-row decomposition of the analog
        precoding matrix which also considers the influence of digital precoding. Then we match each antenna to the optimal RF chain and iterate to obtain the optimal phase. 
        \item[$\bullet$] We evaluate the performance of the proposed alternating optimization algorithms under different structures. Specifically, we validate the proposed algorithms in simulation for FC structures and PC structures separately. The simulation results demonstrate that our proposed wideband hybrid precoding with low-resolution PSs under strong frequency-selective channels in different structures has significant advantages in spectral efficiency. In addition, the benefits of our proposed schemes are further increased under large bandwidth.
\end{enumerate}

\textit{Organization}: The remainder of this paper is organized as follows. In Section \ref{two}, we introduce the system model and channel model, followed by the problem formulation. Then we propose three different alternating optimization algorithms to solve the low-resolution PSs wideband hybrid precoding under strong frequency-selective channels for the three different connection structures in Section \ref{three} and \ref{four}. The simulation results are shown in Section \ref{four}. Finally, we conclude our paper in Section \ref{five}.

\textit{Notations}: In this paper, bold lowercase and bold uppercase letters represent vectors and matrices, respectively. $\mathbb{C}$ denotes the set of complex numbers. $min\{A, B\}$ denotes the minimum between $A$ and $B$. $\|\cdot\|_\mathsf{0}$ is the 0-norm of the vector. $\|\cdot\|_F$ denotes the Frobenius norm of the matrix. $|\cdot|$ represents the value of the determinant. $angle(\cdot)$ is the operation of  computing element phase. $\mathbf{I}_{N}$ denotes an N-dimensional identity matrix. $(\cdot)^H$ demotes the conjugate transposition of the matrix.

\section{System Model and Channel Model}
\label{two}
In this section, we will introduce the system model, the wideband THz channel model, and the problem formulation.

\subsection{System Model}
\label{section2}
In this paper, we consider a wideband THz MIMO-OFDM system, which works at $f_c$ center frequency with $K$ subcarriers and bandwidth $B$. Uniform Linear Array (ULA) are deployed at both the transmitter and the receiver. The transmitter is equipped with $N_t$ antennas and $N^t_{RF}$ RF chains. There are $N_r$ antennas and $N^t_{RF}$ RF chains in the receiver. The signal is transmitted through $N_s$ data streams. Considering the hybrid structure, the number of antennas is much larger than the number of RF chains, which satisfies $N^t_{RF} \ll N_t$ and $N^r_{RF} \ll N_r$. Limited by the number of RF chains, the number of $N_s$ data streams are constrained as $N_s < min\{N^t_{RF}, N^r_{RF}\}$. 

\begin{figure}[htbp]
	\centering
        \captionsetup[subfloat]{font=small,labelfont=rm}
	\subfloat[Fully connected]{\includegraphics[width=.5\columnwidth]{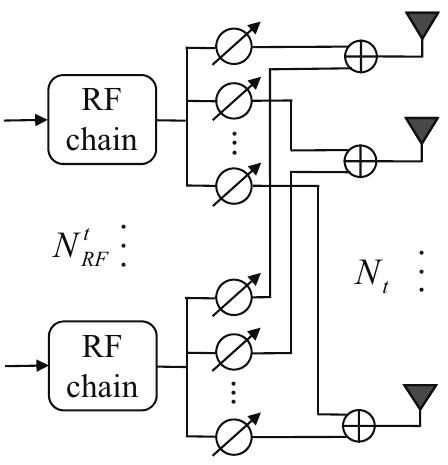} \label{1a}} \hspace{10pt}
	\subfloat[Partially connected]{\includegraphics[width=.4\columnwidth]{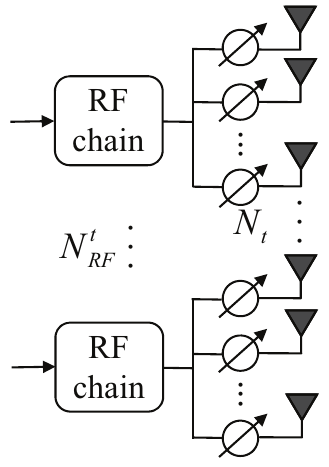} \label{1b}}\\ \hspace{5pt}
	\subfloat[Partially-connected Structure with Dynamic Subarray]{\includegraphics[width=.58\columnwidth]{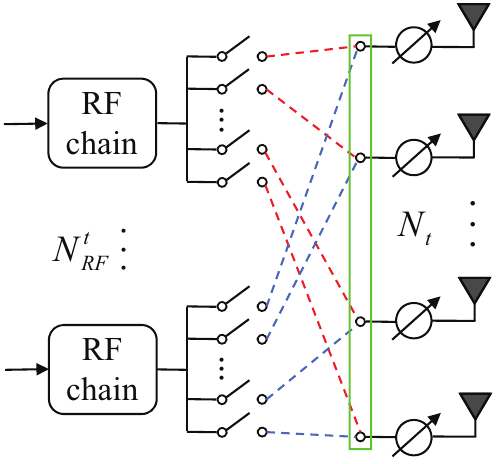} \label{1c}}
	\caption{Three different analog parts of the structure of THz hybrid precoding: each RF chain is connected to all antennas in (a); each RF chain is connected to fixed subarray in (b); each RF chain is connected to dynamic subarray. Notably, in the green box in (c), the nodes have restrictions: each node can connect to any RF chain, with no more than one RF chain at one time.}
    \label{fig1}
\end{figure}

We consider resolving the performance degradation caused by low-resolution PSs and beam squint under three different hybrid structures\textemdash FC structure, PC structure and PC structure with dynamic-subarray, shown in Fig. \ref{fig1}. Each of the three structures has its own advantages. In FC structures, each RF chain connects all antennas, which performs the best among them but requires a large number of PSs and complex connections, shown in Fig. \ref{fig1}\subref{1a}. PC structures reduce the need for the number of PSs, in which each RF chain is connected only to a subset of the antenna array\cite{linx2017hybrid}. Dynamic subarray is a compromise between the FC and PC structures, in which the antennas are adaptively partitioned into several subarrays through switch network and each RF chain maps to one of them \cite{park2017dynamic}, as illustrated in Fig. \ref{fig1}\subref{1c}. The dynamic-subarray increases the degree of freedom of PC structures, with performance closer to FC structures. 

Although the three structures differ in the analog part, the baseband digital precoding is the same.  The
$N_s$ data streams signals are first processed by the baseband digital precoder $\bf{F}_\mathit{BB}\mathit{[k]} \in \mathbb{C}^\mathit{N^t_{RF} \times N_s}$, where $k \in \{1,2,\cdots,K\}$ is the index of the subcarriers. Following the K-dim IFFT and adding CP, the signals are up-convert to RF domain. The following will detail the differences in the system model in the analog part. 
\subsubsection{Analog Precoder for FC Structures} 
In the FC structure, the processed RF signals are transmitted to the PS network and precoded by the analog precoder $\bf{F}_\mathit{RF} \in \mathbb{C}^\mathit{N_t \times N^t_{RF}}$. Different from the baseband digital precoder, $\bf{F}_\mathit{RF}$ is frequency-flat, which means $\bf{F}_\mathit{RF}$ provides the same phase shift for all subcarriers. In addition, the constant modulus (CM) is imposed to each PS according to $[\bf{F}_\mathit{RF}]_\mathit{i,j} = \frac{\mathsf{1}}{\sqrt{\mathit{N_t}}}\mathit{e^{j\vartheta_{i,j}}}$, in which the phase $\vartheta_{i,j}$ is quantized as $\vartheta_{i,j} \in \mathcal{B} \triangleq  \{\frac{2\pi i}{2^b}|i=0,1,2,\cdots,2^b-1\}$. Notably, there are $\mathit{N^tN^t_{RF}}$ PSs in the FC structure, so $\bf{F}_\mathit{RF}$ has no zero element.
\subsubsection{Analog Precoder for PC Structures}
As mentioned above, in the PC structure, each RF chain only connects a subset of the antenna array with the size of $\frac{N_t}{N^t_{RF}}$. The subarray is fixed, and the antennas in the subarray are adjacent. Therefore, the analog precoding matrix satisfies a block diagonal format, which can be formulated as,
\begin{equation}
	\begin{aligned}
		\begin{split}
			\bf{F}_\mathrm{RF} &= \left[ {\begin{array}{*{5}{c}}
					{\mathbf{v}_\mathit{1}} & {\mathbf{0}} & \cdots & {\mathbf{0}}\\
					{\mathbf{0}} & {\mathbf{v}_\mathit{2}} & \cdots & {\mathbf{0}} \\
					\vdots & \vdots & \ddots & \vdots\\
					{\mathbf{0}} & {\mathbf{0}} & \cdots & {\mathbf{v}_\mathit{N^t_{RF}}}
			\end{array}} \right].
		\end{split}
	\end{aligned}
\end{equation}
$\mathbf{v}_\mathit{i} \in \mathbb{C}^{\mathit{\frac{N^t}{N^t_{RF}}} \times 1}$ is the analog precoding matrix of the $i_{th}$ RF chain. In the PC structure, the CM is also imposed to each element in $\mathbf{v}_\mathit{i}$ according to $[\mathbf{v}_\mathit{i}]_{j} = \frac{1}{\sqrt{\mathit{N_t}}}\mathit{e^{j\vartheta_{j}}}$, in which the phase $\vartheta_{j}$ is quantized as $\vartheta_{j} \in \mathcal{B} \triangleq  \{\frac{2\pi i}{2^b}|i=1,2,\cdots,2^b\}$. Obviously, the number of PSs in the partially-connected structure is much less than the number of PSs in the FC structures, with only $N_t$.
\subsubsection{Analog Precoder for Partially-Connected Structures With Dynamic Subarray}
Although the PC structure is energy-efficient compared with the FC structure, it results the severe performance degradation. To compensate the loss, dynamic subarray is introduced in our work. In the PC structure with dynamic subarray, each antenna can select any RF chain to connect with through the switch network (SW). It's feasible to utilize the switch to control the access of antenna because of the low insertion loss (only about 1dB) and high switching speed\cite{7394147}. It should be noticed that the sets of antennas connected to different RF chains have no intersection, which ensure that the number of PSs is $N_t$. As before, we define the analog matrix $\bf{F}_\mathrm{RF} \in \mathbb{C}^\mathit{N_t \times N^t_{RF}}$. The dynamic subarray improves the flexibility between the antennas and the RF chain. Therefore, the analog precoding matrix is no longer limited to the form of a diagonal matrix. Its more relaxed constraints can be represented as the $i_{th}$ row of $\bf{F}_\mathrm{RF}$ satisfying $\|[\bf{F}_\mathrm{RF}]_\textit{i,*}\|_\mathsf{0} \leq \mathsf{1}$, which means only one non-zero element exists in each row. The constant modulus (CM) and low resolution is also imposed to the non-zero element.

\subsection{Channel Model}
Processed by the baseband digital precoder and the analog precoder, the transmitted signal on the $k_{th}$ subcarrier ${\bf x}[k]$ can be expressed as,
\begin{equation}
	\label{signalx}
	{\bf x}[k] = {\bf{F}_\mathrm{RF}}{\bf{F}_\mathrm{BB}}[k]\bf s[\mathit k],
\end{equation} 
where ${\bf s}[k]$ is the original signal conveyed by the $k_{th}$ subcarrier with the size of $N_s \times 1$. The original signal satisfies $\rm E[\bf s[\mathit k]] = \mathsf{0}$ and $\rm E[\bf s[\mathit k]\bf s^\mathit{H}[\mathit k]]=I_\mathit{N_{s}}$. The total power of analog precoder and digital precoder is normalized according to $\|{\bf{F}_\mathrm{RF}}{\bf{F}_\mathrm{BB}}[k]\|_{F}^2 = N_sP_t$, where $P_t$ is the transmission power of each data stream. We assumed the transmission power is equally sharing for all subcarriers. 

Before received, the signal ${\bf x}[k]$ is transmitted over a broadband frequency-selective channel ${\bf H}{\mathit{[k]}}$,$k=1,2,\cdots,K$. Considering the sparsity of the THz channel, the THz channel  at $k_{th}$ subcarrier can be formulated as,
\begin{equation}
	\label{channelh}
	\begin{split}
		\mathbf{H}[k]= \gamma\sum_{m=1}^{N_{\mathit{cl}}} \sum_{n=1}^{N_{\mathit{ray}}} g_{m,n,k} \beta_{m,n,k} \mathbf{a}_{\mathit{r}}\left(\phi^r_{m, n, k}\right) \mathbf{a}_{\mathit{t}}^{H}\left(\varphi^t_{m, n, k}\right),
	\end{split}
\end{equation} 
where $\gamma = \sqrt{\frac{N_tN_r}{N_{cl}N_{ray}}}$ is the normalization factor. Considering the frequency-selective channel, we consider the gain of subcarriers is different. So $g_{m,n,k}$ is the path gain of  the $k_{th}$ subcarrier of the $n_{th}$ ray in the $m_{th}$ cluster. $\beta_{m,n,k} = e^{-j2{\pi}{\tau_{m,n}}f_k}$ is the delay component, in which $\tau_{m,n}$ represents the time delay of the $m_{th}$ ray in the $n_{th}$ cluster. We consider a uniform linear array (ULA) at both the transmitter and the receiver. Therefor, the array response vector $\mathbf{a}_{\mathit{r}}\left(\phi^r_{m, n, k}\right)$ and $\mathbf{a}_{\mathit{t}}\left(\varphi^t_{m, n, k}\right)$ can be formulated as,
\begin{equation}
	\label{ar}
	\begin{split}
		\mathbf{a}_{\mathit{r}}(\phi^r_{m, n, k})=\displaystyle\frac{1}{\sqrt{N_r}}\left[1, e^{j{\pi}\phi^r_{m, n, k}},\cdots,e^{j(N_r-1){\pi}\phi^r_{m, n, k}}\right],
	\end{split}
\end{equation} 
\begin{equation}
	\label{at}
	\begin{split}
		\mathbf{a}_{\mathit{t}}(\varphi^t_{m, n, k})=\displaystyle\frac{1}{\sqrt{N_t}}\left[1, e^{j{\pi}\varphi^t_{m, n, k}},\cdots,e^{j(N_t-1){\pi}\varphi^t_{m, n, k}}\right],
	\end{split}
\end{equation} 
where $\phi_{m,n,k}$ is the normalized angle of departure (AoA) of the $n_{th}$ ray in the $m_{th}$ cluster and $\varphi_{m,n,k}$ is the normalized angle of arrival (AoD) of the $n_{th}$ ray in the $m_{th}$ cluster, which can be formulated as,
\begin{equation}
	\label{phir}
	\begin{split}
		\phi^t_{m, n, k} = \frac{f_k}{f_c}sin\theta_{m, n},
	\end{split}
\end{equation}
\begin{equation}
	\label{varphit}
	\begin{split}
		\varphi^t_{m, n, k} = \frac{f_k}{f_c}sin\vartheta_{m, n}.
	\end{split}
\end{equation}  
$\theta_{m, n}$ and $\vartheta_{m, n}$ represents the actual angle of departure and the actual angle of arrival respectively. $f_k$ is the frequency at the $k_{th}$ subcarrier which can be formulated as $f_k = f_c - \frac{B}{2} + \frac{B}{K}(k-\frac{1}{2})$. \eqref{phir} and \eqref{varphit} are derived based on the distance $d = \frac{c}{2f_c}$ between adjacent antennas, where c is the speed of light. It's obvious that the normalized AoA and AoD are frequency-dependent. However, the analog precoder is frequency-independent. This is the main cause of beam squint. The impact of beam squint on wideband hybrid precoding with low-resolution will be discussed in Section \ref{three}.

\subsection{Problem Formulation}
The purpose of hybrid precoding is to maximize the overall spectral efficiency of the wideband THz communication system under a power spectral density constraint for each subcarrier\cite{9433528}. In this paper, what we mainly focus on is the hybrid precoding at the transmitter. So we assume that the hybrid precoding at the receiver is optimal. Moreover, we assume that the perfect CSI is available at the BS. Therefore, the optimization objective can be expressed by the sum achievable rate as,
\begin{equation}
	\begin{aligned}
		\label{achievablerate}
		R_{\mathrm{avg}}=\frac{1}{K} \sum_{k=1}^{K} \log _{2}\left(\mid \mathbf{I}_{N_{\mathrm{r}}}\right.&+  \frac{P_t}{\sigma_{n}^{2}}\left(\mathbf{H}[k] \mathbf{F}_{\mathrm{RF}} \mathbf{F}_{\mathrm{BB}}[k]\right. \\
		& \left.\left.\times \mathbf{F}_{\mathrm{BB}}^{H}[k] \mathbf{F}_{\mathrm{RF}}^{H} \mathbf{H}^{H}[k]\right) \mid\right),
	\end{aligned}
\end{equation}
where ${\sigma_{n}^{2}}$ denotes the noise power of each subcarrier. There are some differences in constraints due to different structural constraints. The different constraints will be given separately in the Section \ref{three} and Section \ref{four}.

\section{Hybrid Precoding With Low-Resolution PSs For Fully Connected Structures}
\label{three}

In this section, we first analyze the impact of the beam squint on wideband hybrid precoding with low-resolution PSs, which shows that the beam squint is still severe with low-resolution. To maximize the spectral efficiency over frequency selective channels, a heuristic algorithm is proposed in this section, which optimizes the analog precoder and digital precoder alternately. The weight between different subcarriers is fully considered in the proposed scheme, which makes the sum of all subcarriers approximate to the optimum.

\subsection{Beam Squint With Low-resolution PSs}
As well known, the PSs in hybrid precoding structure are frequency flat, which means the 
phase shifting is independent of frequency. However,  in the wideband THz MIMO-OFDM system, the channel is frequency-dependent according to \eqref{phir}\eqref{varphit}. This contradiction leads to beams of different frequencies pointing in different directions. This contradiction leads to beams of different frequencies pointing in different directions, which is called beam squint. The prior analysis of beam squint is always based on the infinite-resolution or high-resolution PSs. In this part, the analysis of the effect of beam squint is with low-resolution PSs. Because of the discrete phase, analog precoding vector obtained through SVD has significant errors and hardly conducts a qualitative analysis. Therefore, we adopt an iterative process to calculate the analog precoding vector and conduct a quantitative analysis\cite{sohrabi2016hybrid}. The iterative process is given in Algorithm 1. 
\begin{algorithm}[!h]
	\caption{Algorithm of Analog Precoding For Narrowband}
	\label{alg:AOS}
	\renewcommand{\algorithmicrequire}{\textbf{Input:}}
	\renewcommand{\algorithmicensure}{\textbf{Output:}}
	
	\begin{algorithmic}[1]
		\REQUIRE $\mathbf{H}_{f_c}$, $P_t$, $\sigma^2$  
		\ENSURE $\mathbf{F}_{\mathrm{RF}}$    
		
		\STATE  Initialize $\mathbf{F}_{\mathrm{RF}}$
		\STATE Calculate $\hat{\textbf{H}}_{f_c}=\textbf{H}_{f_c}^H\textbf{H}_{f_c}$.
  
		\FOR{each $i \in [1,N_{RF}^t]$}
		\STATE Calculate $\textbf{C}_i=\textbf{I}+\frac{P_t}{\sigma^2}({\mathbf{F}}_{\textrm{RF},\backslash i})^H\textbf{H}_{f_c}{\textbf{F}}_{\textrm{RF},\backslash i}$.
		\STATE Calculate $\textbf{G}_{i}\!=\!\frac{P_t}{\sigma^{2}}\hat{\textbf{H}}_{f_c}\!-\!\frac{P_t^2}{\sigma^{4}}\hat{\textbf{H}}_{f_c}{{\textbf{F}}}_{\textrm{RF},\backslash i}\textbf{C}_{i}^{-1}({\textbf{F}}_{\textrm{RF},\backslash i})^{H}\hat{\textbf{H}}_{f_c}$.
		\FOR{each $j \in [1,N_t]$}
		\STATE Calculate $\eta_{ji}=\sum_{\ell\neq i}\textbf{G}_{i}(j,\ell)\textbf{F}_{\textrm{RF}}(\ell,i)$.
            \STATE Calculate $\textbf{F}_{\textrm{RF}}(j,i)=\begin{cases}1,&\text{if }\eta_{ji}=0,\\\frac{\eta_{ji}}{|\eta_{ji}|},&\text{otherwise.}\end{cases}$
		\ENDFOR
		\ENDFOR
		\STATE Check convergence. If not, go back to step 2.
		\RETURN Outputs
	\end{algorithmic}
\end{algorithm}
\begin{figure}
	\centering
	\includegraphics[width=0.5\textwidth]{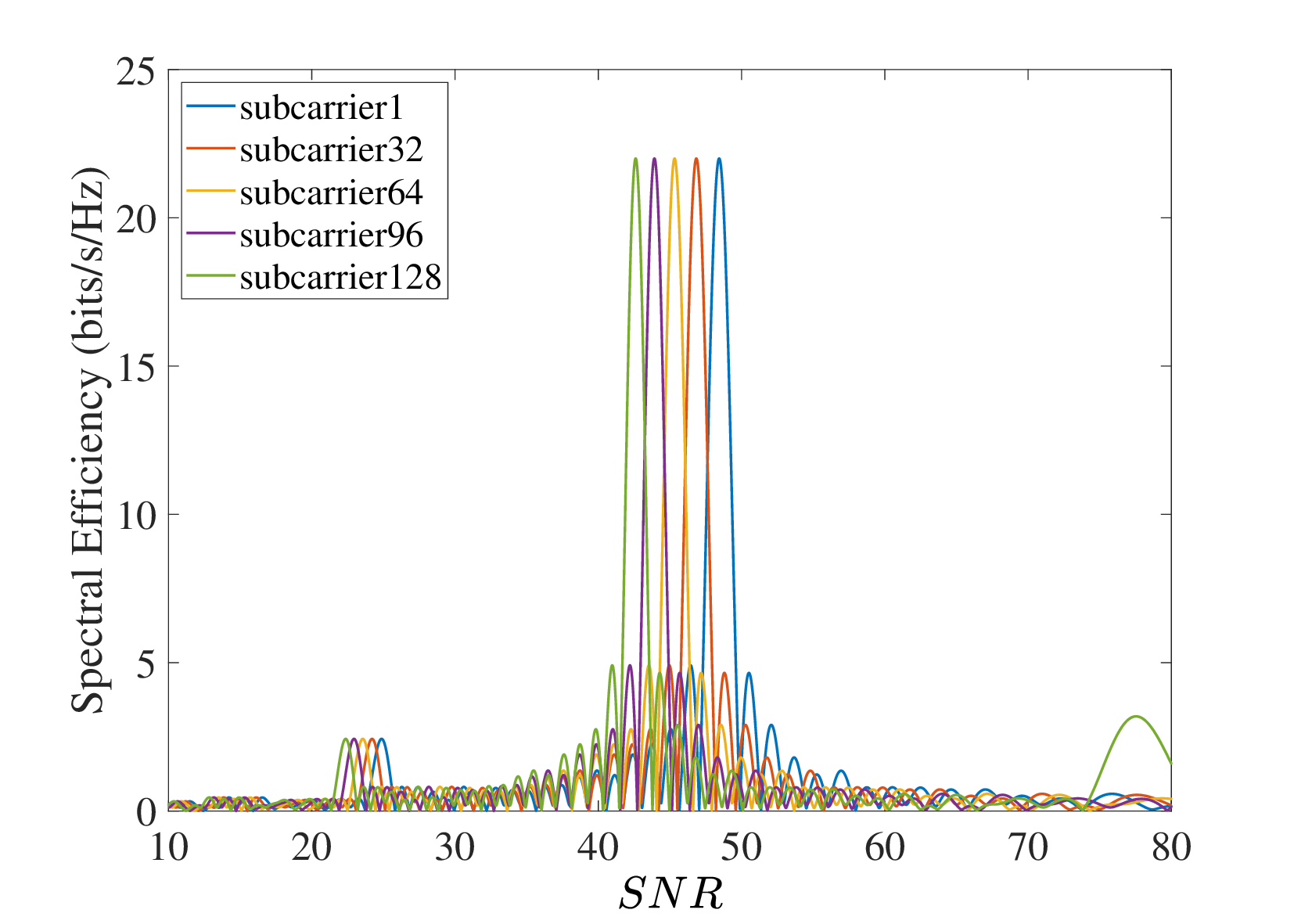}
	\captionsetup{font=small}
	\caption{An example of angular domain gain plot of beam split with $N_t=128$, $N^t_{RF}=1$, $K=128$ and $\theta_u=45^{\circ}$. The resolution of PS is 3bit. The system works at 300GHz center frequency with 30GHz bandwidth. Obviously, at the interest angle $\theta_u=45^{\circ}$, the gain of the subcarriers away from the center frequency drops significantly.}
	\label{beamsplit}
\end{figure}
Fig. \ref{beamsplit} shows an example of angular domain gain plot under the effect of beam split. We can observe that if we design the analog precoding vector directed at $\theta_u = \frac{\pi}{4}$ at center frequency $f_c=300GHz$, the beams at the other frequencies will deviate from the direction $\theta_u = \frac{\pi}{4}$, which demonstrates the severe angular domain gain due to the impact of beam squint. Therefore, it's necessary to take beam squint into consideration in the design of hybrid precoding.

\subsection{Design of Digital Precoder $\mathbf{F}_{\mathrm{BB}}[k]$}
In this part, the optimal digital precoder is presented with the fixed analog precoder. In the FC structure, the optimization objective and constraints can be expressed as,
\begin{equation}
	\begin{aligned}
		\label{achievableratefully}
		R_{\mathrm{avg}}=\frac{1}{K} \sum_{k=1}^{K} \log _{2}&\left(\mid \mathbf{I}_{N_{\mathrm{r}}}\right.+  \frac{P_t}{\sigma_{n}^{2}}\left(\mathbf{H}[k] \mathbf{F}_{\mathrm{RF}} \mathbf{F}_{\mathrm{BB}}[k]\right. \\
		& \left.\left.\times \mathbf{F}_{\mathrm{BB}}^{H}[k] \mathbf{F}_{\mathrm{RF}}^{H} \mathbf{H}^{H}[k]\right) \mid\right),\\
		\text { s.t. }&\left|\left[\mathbf{F}_{\mathrm{RF}}\right]_{i, j}\right|=\frac{1}{\sqrt{N_{\mathrm{t}}}}, \quad \forall i, j, \\
		&\left\|\mathbf{F}_{\mathrm{RF}} \mathbf{F}_{\mathrm{BB}}[k]\right\|_{F}^{2}=N_{\mathrm{s}}P_{\mathrm{t}},\\
		&angle(\left[\mathbf{F}_{\mathrm{RF}}\right]_{i, j}) \in \mathcal{B}.
	\end{aligned}
\end{equation}
With the analog precoder $\mathbf{F}_{\mathrm{RF}}$ fixed, the effective channel of the $k_{th}$ subcarrier can be expressed as,
\begin{equation}
 	\mathbf{H}_\textit{eff}[k] = \mathbf{H}[k]\mathbf{F}_{\mathrm{RF}}(\mathbf{F}_{\mathrm{RF}}^H\mathbf{F}_{\mathrm{RF}})^{-\frac{1}{2}}.
\end{equation}
The digital precoder is implemented by the baseband. In baseband, the design of digital precoder $\mathbf{F}_{\mathrm{BB}}[k]$ of different subcarriers is mutually independent. Therefore, the optimal digital precoder $\mathbf{F}_{\mathrm{BB}}[k]$ is decoupled, which can be formulated as,
\begin{equation}
    \label{digitalprecoding}
	\mathbf{F}_{\mathrm{BB}}[k] = (\mathbf{F}_{\mathrm{RF}}^H\mathbf{F}_{\mathrm{RF}})^{-\frac{1}{2}}\mathbf{V}_{Ns}[k].
\end{equation}
where $\mathbf{V}_{Ns}[k]$ consists the first $N_s$ columns of $\mathbf{V}[k]$. $\mathbf{V}[k]$ is the right singular vectors of $\mathbf{H}_\textit{eff}[k]$. Considering the power constraints in \eqref{achievableratefully}, the digital precoder is normalized according to,
\begin{equation}
    \label{normalization}
	\mathbf{F}_{\mathrm{BB}}[k] = \frac{\sqrt{N_sP_t}}{\left\|\mathbf{F}_{\mathrm{RF}} \mathbf{F}_{\mathrm{BB}}[k]\right\|_{F}}\mathbf{F}_{\mathrm{BB}}[k].
\end{equation}

\subsection{Design of Analog Precoder $\mathbf{F}_{\mathrm{RF}}$} 
The performance of hybrid precoding is influenced by the collective action of all PSs, hence the phase optimization for each PS is interdependent. Limited by the low-resolution, it is difficult to optimize all discrete PSs at the same time through matrix decomposition because the orthogonality between the data stream precoding vectors is broken. Motivated by this, we optimize each PS in turn to decouple analog precoding problem caused by low resolution.

We observe that $\mathbf{F}_{\mathrm{BB}}[k]$ and $\mathbf{F}_{\mathrm{RF}}$ are both in the same item, which restricts the further decomposition of the objective function \eqref{achievableratefully}. So we first transform \eqref{achievableratefully} to \eqref{achievableratefully1} and only show the transformation process of $R[k]$ on each single subcarrier for simplicity. In \eqref{achievableratefully1}, $(a)$ can be easily proofed by Sylvester's determinant theorem which is described as $|\mathbf{I+XY}|=|\mathbf{I+YX}|$. To assure the invertibility of $\mathbf{F}_{\mathrm{BB}}[k]\mathbf{F}_{\mathrm{BB}}^{H}[k]$, we add $\alpha\mathbf{I}_{N_{\mathrm{r}}}$ to $\mathbf{F}_{\mathrm{BB}}[k]\mathbf{F}_{\mathrm{BB}}^{H}[k]$. It is worth nothing that $\alpha$ is a very small scalar so it has little effect on the value of $R[k]$, which makes the equation $(b)$ true. $(c)$ follows because of $|\mathbf{XY}|=|\mathbf{YX}|$ when both $\mathbf{X}$ and $\mathbf{Y}$ are square matrix. Obviously, \eqref{achievableratefully1} successfully separate $\mathbf{F}_{\mathrm{BB}}[k]$ and $\mathbf{F}_{\mathrm{RF}}$, which makes $\mathbf{F}_{\mathrm{BB}}[k]$ related only to the first fraction and the first term of the second fraction, while $\mathbf{F}_{\mathrm{RF}}$ is related only to the second term of the second fraction. In this
paper, we develop an iterative block coordinate descent method to optimize $\mathbf{F}_{\mathrm{RF}}$ and $\mathbf{F}_{\mathrm{BB}}[k]$. The optimal $\mathbf{F}_{\mathrm{BB}}[k]$ can easily calculated according to \eqref{digitalprecoding} and \eqref{normalization} if $\mathbf{F}_{\mathrm{RF}}$ is known. The only remaining issue lies in the calculation of $\mathbf{F}_{\mathrm{RF}}$, which will be given in the following part. 

\begin{figure*}[!hb]
	\begin{equation}
		\begin{split}
			\label{achievableratefully1}
			R[k]=&\log _{2}\left(\middle | \mathbf{I}_{N_{\mathrm{r}}}\!+\!  \frac{P_t}{\sigma_{n}^{2}}\mathbf{H}[k] \mathbf{F}_{\mathrm{RF}} \mathbf{F}_{\mathrm{BB}}[k]\mathbf{F}_{\mathrm{BB}}^{H}[k] \mathbf{F}_{\mathrm{RF}}^{H} \mathbf{H}^{H}[k] \middle |\right)\\
			\overset{(a)}{=}&\log _{2}\left(\middle | \mathbf{I}_{N_{\mathrm{r}}}\!+ \! \frac{P_t}{\sigma_{n}^{2}}(\mathbf{F}_{\mathrm{BB}}[k]\mathbf{F}_{\mathrm{BB}}^{H}[k]+\alpha\mathbf{I}_{N_{RF}^t}) \mathbf{F}_{\mathrm{RF}}^{H} \mathbf{H}^{H}[k]\mathbf{H}[k] \mathbf{F}_{\mathrm{RF}}  \middle |\right)\\
			\overset{(b)}{\approx}&\log _{2}\left(\middle | (\mathbf{F}_{\mathrm{BB}}[k]\mathbf{F}_{\mathrm{BB}}^{H}[k]+\alpha\mathbf{I}_{N_{RF}^t})\{(\mathbf{F}_{\mathrm{BB}}[k]\mathbf{F}_{\mathrm{BB}}^{H}[k]+\alpha\mathbf{I}_{N_{RF}^t})^{-1}\!+ \! \frac{P_t}{\sigma_{n}^{2}} \mathbf{F}_{\mathrm{RF}}^{H} \mathbf{H}^{H}[k]\mathbf{H}[k] \mathbf{F}_{\mathrm{RF}}\}  \middle |\right)\\
			\overset{(c)}{=}&\underbrace{\log _{2}\left(\middle | \mathbf{F}_{\mathrm{BB}}[k]\mathbf{F}_{\mathrm{BB}}^{H}[k]+\alpha\mathbf{I}_{N_{RF}^t}\middle |\right)}_{Fraction 1}+\underbrace{\log _{2}\left(\middle | (\mathbf{F}_{\mathrm{BB}}[k]\mathbf{F}_{\mathrm{BB}}^{H}[k]+\alpha\mathbf{I}_{N_{RF}^t})^{-1}\!+ \! \frac{P_t}{\sigma_{n}^{2}} \mathbf{F}_{\mathrm{RF}}^{H} \mathbf{H}^{H}[k]\mathbf{H}[k] \mathbf{F}_{\mathrm{RF}} \middle|\right)}_{Fraction 2}.
		\end{split}
	\end{equation}
\end{figure*}

Before introducing the calculation of $\mathbf{F}_{\mathrm{RF}}$, we first simplify and clarify the optimization objective \eqref{achievableratefully1}. We define the Hermite matrix $\mathbf{Q}[k]$ as  $\mathbf{F}_{\mathrm{BB}}[k]\mathbf{F}_{\mathrm{BB}}^{H}[k]+\alpha\mathbf{I}_{N_{RF}^t}$. Then \eqref{achievableratefully1} can be further expressed as,
\begin{equation}
	\label{achievableratefully2}
	\begin{split}
		R[k]=\log _{2}\left(\middle |\mathbf{Q}[k]\middle |\right)+\log _{2}\left(\middle | (\mathbf{Q}[k])^{-1}+\right.\\
		\left.\frac{P_t}{\sigma_{n}^{2}} \mathbf{F}_{\mathrm{RF}}^{H} \mathbf{H}^{H}[k]\mathbf{H}[k] \mathbf{F}_{\mathrm{RF}} \middle|\right).
	\end{split}
\end{equation}

So far, we have decoupled $\mathbf{F}_{\mathrm{BB}}[k]$ and $\mathbf{F}_{\mathrm{RF}}$ and transformed $R[k]$ from \eqref{achievableratefully} to \eqref{achievableratefully2}. However, it's still hard to optimize directly in the form of \eqref{achievableratefully2}. Inspired by \cite{sohrabi2017hybrid}, we propose to further transform the optimization objective \eqref{achievableratefully2} by decomposing each column of $\mathbf{F}_{\mathrm{RF}}$. It is worth noting that, unlike \cite{sohrabi2017hybrid}, our goal is to optimize the  achievable rate $R_{\mathrm{avg}}$ itself instead of the upper bound. Therefore, there is a significant difference in the approach. We consider the impact of the $m_{th}$ column of $\mathbf{F}_{\mathrm{RF}}$ on the $R[k]$ and denote $\mathbf{F}_{\mathrm{RF},\backslash m}$ as the analog precoding matrix excluding the $m_{th}$ column $\mathbf{F}_{\mathrm{RF},m}$. Furthermore, we define the Hermite channel $\hat{\mathbf{H}}[k] = \mathbf{H}^{H}[k]\mathbf{H}[k]$ for simplicity. According to properties of determinants, $R[k]$ can be arranged as (\ref{achievableratefully31}--\ref{achievableratefully33}) shown at the bottom, where \eqref{achievableratefully31} is based on the elementary row and column operation of determinant and \eqref{achievableratefully32} is satisfied since $\begin{vmatrix}A&B\\C&D\end{vmatrix}=\begin{vmatrix}D\end{vmatrix}\begin{vmatrix}A-BD^{-1}C\end{vmatrix}$ when $D$ is invertable.
\begin{figure*}[!ht]
	\begin{align}
		\label{achievableratefully31}
		R[k]=&\log _{2}\left(\middle |\mathbf{Q}[k]\middle |\right)\!+\!\log _{2}\left(\middle | 
			\begin{matrix}
			[(\mathbf{Q}[k])^{-1}]_{m,m}\!+\!\frac{P_t}{\sigma_{n}^{2}} \mathbf{F}_{\mathrm{RF},m}^{H}\hat{\mathbf{H}}[k] \mathbf{F}_{\mathrm{RF},m} & [(\mathbf{Q}[k])^{-1}]_{m,\backslash m}\!+\!\frac{P_t}{\sigma_{n}^{2}} \mathbf{F}_{\mathrm{RF},m}^{H} \hat{\mathbf{H}}[k] \mathbf{F}_{\mathrm{RF},\backslash m} \\
			[(\mathbf{Q}[k])^{-1}]_{\backslash m,m}\!+\!\frac{P_t}{\sigma_{n}^{2}} \mathbf{F}_{\mathrm{RF},\backslash m}^{H} \hat{\mathbf{H}}[k] \mathbf{F}_{\mathrm{RF},m} & [(\mathbf{Q}[k])^{-1}]_{\backslash m,\backslash m}\!+\!\frac{P_t}{\sigma_{n}^{2}} \mathbf{F}_{\mathrm{RF},\backslash m}^{H} \hat{\mathbf{H}}[k] \mathbf{F}_{\mathrm{RF},\backslash m}  \\
			\end{matrix} \middle|\right)\\
			&\begin{aligned}
				\label{achievableratefully32}
				=&\log _{2}\left(\middle |[(\mathbf{Q}[k])^{-1}]_{m,m}\!+\!\frac{P_t}{\sigma_{n}^{2}} \mathbf{F}_{\mathrm{RF},m}^{H} \hat{\mathbf{H}}[k] \mathbf{F}_{\mathrm{RF},m}\middle |\right)+\log _{2}\left(\middle |[(\mathbf{Q}[k])^{-1}]_{\backslash m,\backslash m}\!+\!\frac{P_t}{\sigma_{n}^{2}} \mathbf{F}_{\mathrm{RF},\backslash m}^{H} \hat{\mathbf{H}}[k] \mathbf{F}_{\mathrm{RF},\backslash m}\middle |\times \right. \\
				&\left.\middle |[(\mathbf{Q}[k])^{-1}]_{m,m}\!+\!\frac{P_t}{\sigma_{n}^{2}} \mathbf{F}_{\mathrm{RF},m}^{H} \hat{\mathbf{H}}[k] \mathbf{F}_{\mathrm{RF},m} - ([(\mathbf{Q}[k])^{-1}]_{m,\backslash m}\!+\!\frac{P_t}{\sigma_{n}^{2}} \mathbf{F}_{\mathrm{RF},m}^{H} \hat{\mathbf{H}}[k] \mathbf{F}_{\mathrm{RF},\backslash m})\times\right. \\
				&\left.([(\mathbf{Q}[k])^{-1}]_{\backslash m,\backslash m}\!+\!\frac{P_t}{\sigma_{n}^{2}} \mathbf{F}_{\mathrm{RF},\backslash m}^{H} \hat{\mathbf{H}}[k] \mathbf{F}_{\mathrm{RF},\backslash m})^{-1}([(\mathbf{Q}[k])^{-1}]_{\backslash m,m}\!+\!\frac{P_t}{\sigma_{n}^{2}} \mathbf{F}_{\mathrm{RF},\backslash m}^{H} \hat{\mathbf{H}}[k] \mathbf{F}_{\mathrm{RF},m}) \middle |\right)\\
			\end{aligned}\\
			\label{achievableratefully33}
			=&\log_{2}\left(\middle |\mathbf{Q}[k]\middle |\right)\!+\!\log_{2}\left(\middle |\mathbf{C}[k]\middle |\right)\!+\!\log_{2}\left(\middle | \mathbf{D}[k]\!-\!\frac{P_t}{\sigma_{n}^{2}}\mathbf{F}_{\mathrm{RF},m}^{H}{\mathbf{H}_{em}}[k]\!-\!\frac{P_t}{\sigma_{n}^{2}}{\mathbf{H}_{em}^{H}}[k]\mathbf{F}_{\mathrm{RF},m} \!+\!\frac{P_t}{\sigma_{n}^{2}}\mathbf{F}_{\mathrm{RF},m}^{H}{\mathbf{G}^{H}}[k]\mathbf{F}_{\mathrm{RF},m} \middle |\right).
	\end{align}
\end{figure*} 
In this process, we define some intermediate variables, which can formulated as,
\begin{center}
\begin{align}
	\mathbf{C}[k] = [(\mathbf{Q}[k])^{-1}]_{\backslash m,\backslash       m}\!+\!\frac{P_t}{\sigma_{n}^{2}} \mathbf{F}_{\mathrm{RF},\backslash m}^{H} \hat{\mathbf{H}}[k] \mathbf{F}_{\mathrm{RF},\backslash m}, \label{intermediatevariables:C}\\
        {\mathbf{H}_{em}}[k] = \hat{\mathbf{H}}[k]\mathbf{F}_{\mathrm{RF},\backslash m}(\mathbf{C}[k])^{-1}[(\mathbf{Q}[k])^{-1}]_{\backslash m,m},\label{intermediatevariables:Hem}
\end{align}
\end{center}
\begin{center}
\begin{align}
        &{\mathbf{G}}[k] = \hat{\mathbf{H}}[k]-\frac{P_t}{\sigma_{n}^{2}}\hat{\mathbf{H}}[k]\mathbf{F}_{\mathrm{RF},\backslash m}(\mathbf{C}[k])^{-1}\mathbf{F}_{\mathrm{RF},\backslash m}^H\hat{\mathbf{H}}[k]\label{intermediatevariables:G}\\
        &\begin{aligned}
            \mathbf{D}[k] = [(\mathbf{Q}[k])^{-1}]_{m,m}-[(\mathbf{Q}[k])^{-1}]_{m,\backslash m}(\mathbf{C}[k])^{-1}\times\\
            [(\mathbf{Q}[k])^{-1}]_{\backslash m,m}.
            \label{intermediatevariables:D}
        \end{aligned}
\end{align}
\end{center}

According to \eqref{achievableratefully33}, we observe that the optimization objective can be decomposed into $m$ identical subproblems, namely, finding the optimal $\mathbf{F}_{\mathrm{RF},m}$  with $\mathbf{F}_{\mathrm{RF},\backslash m}$ and $\mathbf{F}_{\mathrm{BB}}$ fixed. $\mathbf{F}_{\mathrm{RF},m}$ is only related to the third term. So the new optimization objective can be rewritten as, 
\begin{equation}
  \begin{aligned}	
    \label{calculationFRF}
    \{\mathbf{F}_{\mathrm{RF},m}^{\star}\}=&\arg\max\frac{1}{K} \sum_{k=1}^{K}\mathbf{M}[k]\\
    &\mathrm{s.t.}\quad\mathbf{F}_{\mathrm{RF},m}(n)\in\mathcal{B},n=1,\ldots,N_{t}.
\end{aligned}  
\end{equation}	
\begin{equation}
    \begin{aligned}
        \mathbf{M}[k]\!=\!\log_{2}\left(\middle | \mathbf{D}[k]\!-\!\frac{P_t}{\sigma_{n}^{2}}\mathbf{F}_{\mathrm{RF},m}^{H}{\mathbf{H}_{em}}[k]\!-\!\frac{P_t}{\sigma_{n}^{2}}{\mathbf{H}_{em}^{H}}[k]\times\right.\\
        \mathbf{F}_{\mathrm{RF},m}\!+\!\left.\frac{P_t}{\sigma_{n}^{2}}\mathbf{F}_{\mathrm{RF},m}^{H}{\mathbf{G}^{H}}[k]\mathbf{F}_{\mathrm{RF},m} \middle |\right)
        \label{intermediatevariables:M}
    \end{aligned}
\end{equation}	
Unfortunately, the $m$ subproblems are not independent, which motivates us to propose an iteration coordinate descent method for phase optimization to update $\mathbf{F}_{RF,m}$ one by one until convergence. Following the guidance, we will present the update scheme for each $\mathbf{F}_{\mathrm{RF},m}$ in each iteration in the following part.

Exhaustively searching for the optimal phase of each PS in $\mathbf{F}_{RF,m}$ would result in extremely high complexity. So we propose an iteration phase optimization method to update the phase of each PS in turn. Specifically, we first consider removing the finite resolution constraint, yielding an approximately optimal phase under infinite resolution conditions. With the other $n-1$ elements fixed, the continuous optimal phase of the $n_{th}$ element of $\mathbf{F}_{\mathrm{RF},m}$ can be written as \eqref{vartheta_1}, which is proved in Appendix \ref{appendix1}.
\begin{equation}
        \label{vartheta_1}
        \hat\vartheta_{n,m} = angle\left(\sum_{k=1}^{K} \frac{[{\mathbf{G}}[k]]_{\backslash n,n}^{H}\mathbf{F}_{\mathrm{RF},m,\backslash n}-{\mathbf{H}_{em,n}}[k]}{\mathbf{M}[k]}\right)
\end{equation}
where $[{\mathbf{G}}[k]]_{\backslash n,n}$ represents the $n_{th}$ column of ${\mathbf{G}}[k]$, excluding the $n_{th}$ row. $\mathbf{F}_{\mathrm{RF},m,\backslash n}$ represents the ($N_t-1$)-dimension vector formed by removing the $n_{th}$ element from $\mathbf{F}_{\mathrm{RF},m}$. Subsequently, because of finite resolution, the optimal phase is matched to the nearest discrete phase in $\mathcal{B}$ in the angular domain, which can be formulated as,
\begin{equation}
    \label{vartheta_2}
    \tilde\vartheta_{n,m} = \mathop{\arg\min}\limits_{\bar\vartheta_{n,m}\in\{\mathcal{B},2\pi\}}|\bar\vartheta_{n,m}-\hat\vartheta_{n,m}|
\end{equation}
Notably, the phase has a period of $2\pi$, which means it's necessary to take the boundary value $2\pi$ into account in \eqref{vartheta_2}. Then, the quantization operation can be formulated as,
\begin{equation}
    \label{vartheta_3}
    \vartheta_{n,m} = \frac{\tilde\vartheta_{n,m}}{2\pi}.
\end{equation}
By exhaustively optimizing each element in $\mathbf{F}_{\mathrm{RF},m}$ according to \eqref{vartheta_1} and \eqref{vartheta_2}, we complete one round of iterative optimization of $\mathbf{F}_{\mathrm{RF},m}$. Notably, the phase calculating of $\mathbf{F}_{\mathrm{RF}}$ is related to the value of $\mathbf{F}_{\mathrm{BB}}$. To ensure the convergence speed of the iteration,  we update $\mathbf{F}_{\mathrm{BB}}$ after the update of each $\mathbf{F}_{\mathrm{RF},m}$, rather than after the completion of all $\mathbf{F}_{\mathrm{RF},m}$ update.\footnote{The calculation of $\mathbf{F}_{\mathrm{BB}}$ involves a certain complexity. Too frequent updates, such as after each PS update, would make each iteration round too slow; too infrequent updates, such as updating after the entire $\mathbf{F}_{\mathrm{RF}}$ update is completed, would slow down the convergence speed.} Thus far, the digital and analog precoding iteration block coordinate algorithm is completed and summarized as Algorithm \ref{alg_fully}.
\begin{algorithm}[!h]
	\caption{Algorithm of Analog Precoding For FC Structure}
	\label{alg_fully}
	\renewcommand{\algorithmicrequire}{\textbf{Input:}}
	\renewcommand{\algorithmicensure}{\textbf{Output:}}
	
	\begin{algorithmic}[1]
		\REQUIRE $\mathbf{H}[k]$, $P_t$, $\sigma^2$  
		\ENSURE $\mathbf{F}_{\mathrm{RF}}$,$\mathbf{F}_{\mathrm{BB}}[k]$ for each $k \in [1,K]$.   
		
		\STATE  Initialize $\mathbf{F}_{\mathrm{RF}}$,$\mathbf{F}_{\mathrm{BB}}[k]$ for each $k \in [1,K]$. 
            \STATE  Calculate $\hat{\textbf{H}}[k]=\textbf{H}[k]^H\textbf{H}[k]$ for each $k \in [1,K]$.
		
		\FOR{each $i \in [1,N_{RF}^t]$}
		\STATE Calculate ${\mathbf{H}_{em}}[k]$, $\mathbf{G}[k]$, $\mathbf{M}[k]$ according to \eqref{intermediatevariables:Hem},\eqref{intermediatevariables:G} and \eqref{intermediatevariables:M}. 
		\FOR{each $j \in [1,N_t]$}
		\STATE Calculate $\hat\vartheta_{j,i}$ and quantize to discrete phases $\vartheta_{j,i}$ according to \eqref{vartheta_1}, \eqref{vartheta_2} and \eqref{vartheta_3}.
		\ENDFOR
            \STATE  Calculate $\mathbf{F}_{\mathrm{BB}}[k]$ for each $k \in [1,K]$ according to \eqref{digitalprecoding} and \eqref{normalization}.
		\ENDFOR
		\STATE Check convergence. If not, go back to step 2.
		\RETURN $\mathbf{F}_{\mathrm{RF}}$,$\mathbf{F}_{\mathrm{BB}}[k]$ for each $k \in [1,K]$.
	\end{algorithmic}
\end{algorithm}


\section{Hybrid Precoding With Low-Resolution PSs For PC Structures}
\label{four}
In section \ref{three}, we introduce the iterative block coordinate descent method to alternately optimize $\mathbf{F}_{\mathrm{RF}}$ and $\mathbf{F}_{\mathrm{BB}}[k]$. In this section, we further extend the proposed iterative alternating optimization algorithm to PC structures. Moreover, we also consider a new dynamic array structure, which enhances the flexibility of PC structures. Based on this, we develop an iterative alternating optimization algorithm motivated by \cite{yan2022dynamic}. 

\subsection{Hybrid precoding Design for Partially Connected Structures}
As each RF chain is only connected to a fixed and non-overlapping subarray in PC structure, the elements that need to optimize are only located on the block diagonal. Therefore, compared with the FC structure, the complexity of the PC structure is reduced a lot because the number of PSs is only $\frac{1}{N_t}$ of the FC structure. The optimization objective of the PC structure can be formulated as, 
\begin{equation}
	\begin{aligned}
		\label{achievableratepartially}
		R_{\text{avg}}=\frac{1}{K} &\sum_{k=1}^{K} \log_2\left(\mid \mathbf{I}_{N_{\mathrm{r}}}\right.+  \frac{P_t}{\sigma_{n}^{2}}\left(\mathbf{H}[k] \mathbf{F}_{\mathrm{RF}} \mathbf{F}_{\mathrm{BB}}[k]\right. \\
		& \left.\left.\times \mathbf{F}_{\mathrm{BB}}^{H}[k] \mathbf{F}_{\mathrm{RF}}^{H} \mathbf{H}^{H}[k]\right) \mid\right),\\
		\text { s.t. }&\left|\left[\mathbf{v}_\mathit{i}\right]_j\right|=\frac{1}{\sqrt{N_{\mathrm{t}}}}, \quad i \in [1,N^t_{RF}], j \in [1,\frac{N_t}{N^t_{RF}}], \\
		&\left\|\mathbf{F}_{\mathrm{RF}} \mathbf{F}_{\mathrm{BB}}[k]\right\|_{F}^{2}=N_{\mathrm{s}}P_{\mathrm{t}},\\
		&angle(\left[\mathbf{v}_\mathit{i}\right]_j) \in \mathcal{B} .
	\end{aligned}
\end{equation}
It's obvious that the difference between \eqref{achievableratefully} and \eqref{achievableratepartially} is the number of PSs to be optimized is different. So we can readily migrate the algorithm designed for FC structures in section \ref{three} to PC structures. The construction of initial $\mathbf{F}_{\mathrm{RF}}$ in the PC structures is given by,
\begin{align}
        \label{initialfrf1}
        &\mathbf{H}_\text{avg} = \frac{1}{K}\sum_{k=1}^{K}\mathbf{H}[k],\\
        \label{initialfrf2}
	[:,:,\mathbf{V}_i] = \mathbf{SVD}&\left\{\mathbf{H}_\text{avg}\left(:,\frac{(i-1)N_t+1}{N^t_{RF}}:\frac{iN_t}{N^t_{RF}}\right)\right\},\\
        \label{initialfrf3}
        &\mathbf{v}_\mathit{i}^{(0)} = \mathbf{V}_i(:,1:N^t_{RF}).
\end{align}
where $i=1,2,\cdots,N^t_{RF}$ and $\mathbf{H}_\text{avg}$ is the average channel matrix. As $\mathbf{F}_{\mathrm{RF}}$ is composed of $\mathbf{v}_\mathit{i}$, it's straightforward to get $\mathbf{F}_{\mathrm{RF}}$ from $\mathbf{v}_\mathit{i}$. The iteration alternating optimization algorithm of the PC structure is summarized in Algorithm \ref{alg_partially}.
\begin{algorithm}[!h]
	\caption{Algorithm of Hybrid Precoding For Partially Connected Structures}
	\label{alg_partially}
	\renewcommand{\algorithmicrequire}{\textbf{Input:}}
	\renewcommand{\algorithmicensure}{\textbf{Output:}}
	
	\begin{algorithmic}[1]
		\REQUIRE $\mathbf{H}[k]$, $P_t$, $\sigma^2$  
		\ENSURE $\mathbf{v}_\mathit{i}$ for each $i \in [1,N^t_{RF}]$,$\mathbf{F}_{\mathrm{BB}}[k]$ for each $k \in [1,K]$ .   
		
		\STATE  Initialize $\mathbf{v}_\mathit{i}$ for each $i \in [1,N^t_{RF}]$ and $\mathbf{F}_{\mathrm{BB}}[k]$ for each $k \in [1,K]$ according to \eqref{normalization} and \eqref{initialfrf3}. 
            \STATE  Calculate $\hat{\textbf{H}}[k]=\textbf{H}[k]^H\textbf{H}[k]$ for each $k \in [1,K]$.
		
		\FOR{each $i \in [1,N_{RF}^t]$}
		\STATE Calculate ${\mathbf{H}_{em}}[k]$, $\mathbf{G}[k]$, $\mathbf{M}[k]$ according to \eqref{intermediatevariables:Hem},\eqref{intermediatevariables:G} and \eqref{intermediatevariables:M}. 
		\FOR{each $j \in [1,\frac{N_t}{N^t_{RF}}]$}
		\STATE Calculate $\hat\vartheta_{j,i}$ and quantize to discrete phases $\vartheta_{j,i}$ according to \eqref{vartheta_1}, \eqref{vartheta_2} and \eqref{vartheta_3}.
            \STATE Let $\mathbf{v}_\mathit{i,j}=\vartheta_{j,i}$. 
		\ENDFOR
            \STATE Update the $i_{th}$ column of $\mathbf{F}_{\mathrm{RF}}$ utilizing $\mathbf{v}_\mathit{i}$. 
            \STATE  Calculate $\mathbf{F}_{\mathrm{BB}}[k]$ for each $k \in [1,K]$ according to \eqref{digitalprecoding} and \eqref{normalization}.
		\ENDFOR
		\STATE Check convergence. If not, go back to step 2.
		\RETURN $\mathbf{F}_{\mathrm{RF}}$ and $\mathbf{F}_{\mathrm{BB}}[k]$ for each $k \in [1,K]$.
	\end{algorithmic}
\end{algorithm}

\subsection{Hybrid precoding Design For Partially Connected Structures With Dynamic Subarray}
Although dynamic subarray increases the flexibility of the PC structure, the optimization difficulty of the analog precoding matrix is simultaneously increased. $\mathbf{F}_{\mathrm{RF}}$  is no longer constrained to be a block diagonal matrix. Therefore, Algorithm \ref{alg_partially} is not suitable in the PC structure with dynamic subarray. The new constraints and optimization objective are given by,
\begin{equation}
	\begin{aligned}
		\label{achievablerateds}
		R_{\mathrm{avg}}=\frac{1}{K} \sum_{k=1}^{K} \log _{2}&\left(\mid \mathbf{I}_{N_{\mathrm{r}}}\right.+  \frac{P_t}{\sigma_{n}^{2}}\left(\mathbf{H}[k] \mathbf{F}_{\mathrm{RF}} \mathbf{F}_{\mathrm{BB}}[k]\right. \\
		& \left.\left.\times \mathbf{F}_{\mathrm{BB}}^{H}[k] \mathbf{F}_{\mathrm{RF}}^{H} \mathbf{H}^{H}[k]\right) \mid\right),\\
		\text { s.t. }&\left|\left[\mathbf{F}_{\mathrm{RF}}\right]_{i, j}\right|\in\left\{\frac{1}{\sqrt{N_{\mathrm{t}}}},0\right\}, \quad \forall i, j, \\
            &\|[\bf{F}_\mathrm{RF}]_\textit{i,*}\|_\mathsf{0} \leq \mathsf{1}, \\
		&\left\|\mathbf{F}_{\mathrm{RF}} \mathbf{F}_{\mathrm{BB}}[k]\right\|_{F}^{2}=N_{\mathrm{s}}P_{\mathrm{t}},\\
		&angle(\left[\mathbf{F}_{\mathrm{RF}}\right]_{i, j}) \in \mathcal{B}.
	\end{aligned}
\end{equation}
As the dynamic subarray is connected by the switch network, the amplitude constraint in \eqref{achievableratefully} is looser than in \eqref{achievableratepartially}. It's allowed that the elements in $\mathbf{F}_{\mathrm{RF}}$ are limited to 0, which means all switches connected to this node are off. To further simplify \eqref{achievablerateds}, we reexpress the analog precoding matrix as,
\begin{equation}
        \label{Frfdy}
	\begin{aligned}
		\begin{split}
			\bf{F}_\mathrm{RF} &= \bf{S}\bf{F}_\mathrm{D}.
		\end{split}
	\end{aligned}
\end{equation}
where $\bf{S}$ is selection matrix with the size of $\mathbb{C}^{\mathit{N_t}\times\mathit{N^t_{RF}2^{bits}}}$, in which the value of the elements in $\bf{S}$ is binary. $\bf{F}_\mathrm{D}\in\mathbb{C}^{\mathit{N^t_{RF}2^{bits}}\times\mathit{N^t_{RF}}}$ is the dictionary matrix used to select the appropriate phase and RF chain, which can be written as,
\begin{equation}
	\begin{aligned}
		\begin{split}
			\bf{F}_\mathrm{D} &= \left[ {\begin{array}{*{5}{c}}
					{\mathbf{f}_\mathrm{D}} & {\mathbf{0}} & \cdots & {\mathbf{0}}\\
					{\mathbf{0}} & {\mathbf{f}_\mathrm{D}} & \cdots & {\mathbf{0}} \\
					\vdots & \vdots & \ddots & \vdots\\
					{\mathbf{0}} & {\mathbf{0}} & \cdots & {\mathbf{f}_\mathrm{D}}
			\end{array}} \right].
		\end{split}
	\end{aligned}
\end{equation}
where $\mathbf{f}_\mathrm{D}=[e^{j\mathcal{B}(1)},e^{j\mathcal{B}(2)},\cdots,e^{j\mathcal{B}(2^{bits})}]$ represents the phase vector composed of elements from $\mathcal{B}$. Then \eqref{achievablerateds} can be rewritten as,
\begin{equation}
	\begin{aligned}
		\label{achievablerateds}
		R_{\mathrm{avg}}=\frac{1}{K} \sum_{k=1}^{K} log _{2}&\left(\mid \right.\mathbf{I}_{N_{\mathrm{r}}}+  \frac{P_t}{\sigma_{n}^{2}}\left(\mathbf{H}[k] \mathbf{S}\mathbf{F}_\mathrm{D} \mathbf{F}_{\mathrm{BB}}[k]\right. \\
		& \left.\left.\times \mathbf{F}_{\mathrm{BB}}^{H}[k] \mathbf{F}_\mathrm{D}^{H}\mathbf{S}^{H} \mathbf{H}^{H}[k]\right) \mid\right),\\
		\text { s.t. }&\|\bf{S}_\textit{i,*}\|_\mathsf{0} \leq \mathsf{1}, \\
            &\bf{S}_\textit{i,j} \in \left\{\mathsf{0,1}\right\},\forall \textit{i}, \textit{j}, \\
		&\left\|\mathbf{S}\mathbf{F}_\mathrm{D}\mathbf{F}_{\mathrm{BB}}[k]\right\|_{F}^{2}=N_{\mathrm{s}}P_{\mathrm{t}}.
	\end{aligned}
\end{equation}
$\mathbf{F}_\mathrm{D}$ is fixed so the optimization objective in analog precoding is transformed from $\mathbf{F}_{\mathrm{RF}}$ to $\mathbf{S}$. Similar to the processing in \eqref{achievableratefully1}, the decoupled optimization objective can be formulated as,
\begin{equation}
	\begin{aligned}
		\label{achievablerateds1}
		R[k]=\log _{2}&\left(\middle | \mathbf{F}_{\mathrm{BB}}[k]\mathbf{F}_{\mathrm{BB}}^{H}[k]+\alpha\mathbf{I}_{N_{RF}^t}\middle |\right)+\\
        \log _{2}&\left(\middle | (\mathbf{F}_{\mathrm{BB}}[k]\mathbf{F}_{\mathrm{BB}}^{H}[k]+\alpha\mathbf{I}_{N_{RF}^t})^{-1}+\right.\\
        &\left. \frac{P_t}{\sigma_{n}^{2}} \mathbf{F}_\mathrm{D}^H\mathbf{S}^{H} \mathbf{H}^{H}[k]\mathbf{H}[k] \mathbf{S}\mathbf{F}_\mathrm{D} \middle|\right).
	\end{aligned}
\end{equation}

The main difficulty in optimizing $\mathbf{S}$ is the L0 norm constraint. Inspired by \cite{yan2022dynamic}, we design a row-by-row decomposition approach to decompose \eqref{achievableratepartially} into $N_t$ sub-problems. The process of decomposition can be formulated as,
\begin{align}
        &\begin{aligned}
        \label{achievablerateds1:1}
            R[k]=&\log_{2}\left(\middle | \mathbf{Q}_\mathrm{DS}[k]\middle |\right)+\log _{2}\left(\middle | (\mathbf{Q}_\mathrm{DS}[k])^{-1}+\frac{P_t}{\sigma_{n}^{2}}\times \right.\\
        &\mathbf{F}_\mathrm{D}^H\left. \begin{bmatrix}
\mathbf{S}_{1:N_t-1,*}\\
\mathbf{S}_{N_t,*}
\end{bmatrix}^{H} \mathbf{H}^{H}[k]\mathbf{H}[k] \begin{bmatrix}
\mathbf{S}_{1:N_t-1,*}\\
\mathbf{S}_{N_t,*}
\end{bmatrix} \mathbf{F}_\mathrm{D}\middle|\right)\\
        \end{aligned}\\
        &\begin{aligned}
        \label{achievablerateds1:2}
            =&\log_{2}\left(\middle | \mathbf{Q}_\mathrm{DS}[k]\middle |\right)\!+\!\log _{2}\left(\middle | (\mathbf{Q}_\mathrm{DS}[k])^{-1}\!+\!\mathbf{F}_\mathrm{D}^H\mathbf{S}_{1:N_t-1,*}^H\times\!\right.\\
        &\hat{\mathbf{H}}_{1:N_t-1,1:N_t-1}\mathbf{S}_{1:N_t-1,*}\mathbf{F}_\mathrm{D}\!+\!\mathbf{F}_\mathrm{D}^H\mathbf{S}_{N_t,*}^H\hat{\mathbf{H}}_{N_t,N_t}\times\\
        &\mathbf{S}_{N_t,*}\mathbf{F}_\mathrm{D}\!+\!\mathbf{F}_\mathrm{D}^H\mathbf{S}_{1:N_t-1,*}^H\hat{\mathbf{H}}_{1:N_t-1,N_t}\mathbf{S}_{N_t,*}\mathbf{F}_\mathrm{D}\!+\!\mathbf{F}_\mathrm{D}^H\times\\
        &\left.\mathbf{S}_{N_t,*}^H\hat{\mathbf{H}}_{N_t,1:N_t-1}\mathbf{S}_{1:N_t-1,*}\mathbf{F}_\mathrm{D}
        \middle|\right)\\
        \end{aligned}
\end{align}
\begin{align}
        &\begin{aligned}
        \label{achievablerateds1:3}
            =&\log_{2}\left(\middle | \mathbf{Q}_\mathrm{DS}[k]\middle |\right)\!+\!\log_{2}\left(\middle | (\mathbf{Q}_\mathrm{DS}[k])^{-1}\!+\!\mathbf{F}_\mathrm{D}^H\mathbf{S}_{1:N_t-1,*}^H\right.\\
            &\left.\hat{\mathbf{H}}_{1:N_t-1,1:N_t-1}\mathbf{S}_{1:N_t-1,*}\mathbf{F}_\mathrm{D}\middle|\right)\!+\!\log_{2}\left(\middle | \mathbf{I}+(\mathbf{D}_{i}[k])^{-1}\!\times\right.\\
        &\mathbf{F}_\mathrm{D}^H(\mathbf{S}_{N_t,*}^H\hat{\mathbf{H}}_{N_t,N_t}\mathbf{S}_{N_t,*}+\mathbf{S}_{1:N_t-1,*}^H\hat{\mathbf{H}}_{1:N_t-1,N_t}\mathbf{S}_{N_t,*}+\\
        &\left.\mathbf{S}_{N_t,*}^H\hat{\mathbf{H}}_{N_t,1:N_t-1}\mathbf{S}_{1:N_t-1,*})\mathbf{F}_\mathrm{D}\middle |\right)\\
        \end{aligned}\\
        &\begin{aligned}
        \label{achievablerateds1:4}
            =&\log_{2}\left(\middle | \mathbf{Q}_\mathrm{DS}[k]\middle |\right)\!+\!\sum_{i=1}^{N_t}\log_{2}\left(\middle | \mathbf{I}+(\mathbf{D}_{i}[k])^{-1}\mathbf{F}_\mathrm{D}^H\!\times\right.\\
&(\mathbf{S}_{N_t,*}^H\hat{\mathbf{H}}_{N_t,N_t}\mathbf{S}_{N_t,*}+\mathbf{S}_{1:N_t-1,*}^H\hat{\mathbf{H}}_{1:N_t-1,N_t}\mathbf{S}_{N_t,*}+\\
        &\left.\mathbf{S}_{N_t,*}^H\hat{\mathbf{H}}_{N_t,1:N_t-1}\mathbf{S}_{1:N_t-1,*})\mathbf{F}_\mathrm{D}\middle |\right)
        \end{aligned}
\end{align}
where $\mathbf{Q}_\mathrm{DS}[k]$ and $\mathbf{D}_{i}[k]$ is the intermediate variable for simplifying the expression of \eqref{achievablerateds1}, which can be expressed as,
\begin{center}
\begin{align}
        &\mathbf{Q}_\mathrm{DS}[k] = \mathbf{F}_{\mathrm{BB}}[k]\mathbf{F}_{\mathrm{BB}}^{H}[k]+\alpha\mathbf{I}_{N_{RF}^t}\label{intermediatevariables:Qds}\\
        \mathbf{D}_{i}[k]& = (\mathbf{Q}_\mathrm{DS}[k])^{-1}\!+\!\mathbf{F}_\mathrm{D}^H\mathbf{S}_{1:i-1,*}^H\hat{\mathbf{H}}_{1:i-1,1:i-1}\mathbf{S}_{1:i-1,*}\mathbf{F}_\mathrm{D}\label{intermediatevariables:Dds}
\end{align}
\end{center}
From \eqref{achievablerateds1:1} to \eqref{achievablerateds1:2}, the $n_{th}$ row is separated from the entire $\mathbf{S}$. \eqref{achievablerateds1:3} holds because $\log_{2}(|\mathbf{X}+\mathbf{Y}|)=\log_{2}(|\mathbf{X}|)+\log_{2}(|\mathbf{I}+\mathbf{X}^{-1}\mathbf{Y}|)$. It's obvious that $\mathbf{S}_{1:N_t-1,*}$ can be further decomposed according to \eqref{achievablerateds1:2}, eventually being broken down into $N_t-1$ terms. We can observe that the optimization of the $n_{th}$ row is only about the first $(n-1)$ rows. Therefore, the optimization can be performed from the first row and proceed row by row.

With $\mathbf{F}_{\mathrm{BB}}$ fixed, the optimization of $\mathbf{F}_{\mathrm{RF}}$ can be decomposed into $N_t$ sub-problems, which can be formulated as, 
\begin{equation}
  \begin{aligned}	
    \label{calculationS}
    \{\mathbf{S}_{i,*}^{\star}\}=\arg\max&\frac{1}{K}\sum_{k=1}^K\log_{2}\left(\middle | \mathbf{E}_{i}[k]\middle |\right),\\
        \text { s.t. }&\|\bf{S}_\textit{i,*}\|_\mathsf{0} \leq \mathsf{1}, \\
            &\bf{S}_\textit{i,j} \in \left\{\mathsf{0,1}\right\},\forall \textit{i}, \textit{j}, \\
    \end{aligned}  
\end{equation}
where $\mathbf{E}_{i}[k]$ is the intermediate variable to simplify \eqref{calculationS}, which can be formulated as,
\begin{equation}
    \begin{aligned}	
    \label{Eik}
        \mathbf{E}_{i}[k]=\mathbf{I}\!+\!(\mathbf{D}_{i}[k])^{(-1)}\mathbf{F}_\mathrm{D}^H(\mathbf{S}_{i,*}^H\hat{\mathbf{H}}_{i,i}\mathbf{S}_{i,*}+\mathbf{S}_{1:i-1,*}^H\hat{\mathbf{H}}_{1:i-1,i}\times\\
        \mathbf{S}_{i,*}+\mathbf{S}_{i,*}^H\hat{\mathbf{H}}_{i,1:i-1}\mathbf{S}_{1:i-1,*})\mathbf{F}_\mathrm{D}.
    \end{aligned}  
\end{equation}
As analyzed in the last paragraph, the $N_t$ sub-problems need to be optimized one by one in ascending order of $i$. To maximize \eqref{calculationS}, we need to find the positions of row vector $\hat{\mathbf{H}}_{i,1:i-1}\mathbf{S}_{1:i-1,*}$ and column vector $\mathbf{S}_{1:i-1,*}^H\hat{\mathbf{H}}_{1:i-1,i}$, which corresponds to the value of $\mathbf{S}_{i,*}$. When the 0 norm of $\mathbf{S}_{i,*}$ is 1, $\mathbf{S}_{i,*}$ has $N^t_{RF}\times 2^b$ possible values, where $b$ is always small. When the 0 norm of $\mathbf{S}_{i,*}$ is 0, $\mathbf{S}_{i,*}$ can only be the 0 vector. Therefore, the optimal $\mathbf{S}_{i,*}$ be solved by a one-dimensional search method which is due to the finite bits of the PSs. The number of searches is $N^t_{RF}\!\times\!2^b\!+\!1$. The optimization algorithm is summarized in Algorithm \ref{alg_dynamic}.



\begin{algorithm}[!h]
	\caption{Algorithm of Hybrid Precoding For Partially Connected Structures}
	\label{alg_dynamic}
	\renewcommand{\algorithmicrequire}{\textbf{Input:}}
	\renewcommand{\algorithmicensure}{\textbf{Output:}}
	
	\begin{algorithmic}[1]
		\REQUIRE $\mathbf{H}[k]$, $P_t$, $\sigma^2$  
		\ENSURE $\mathbf{F}_{\mathrm{RF}}$ and $\mathbf{F}_{\mathrm{BB}}[k]$ for each $k \in [1,K]$.   
		
		\STATE  Initialize $\mathbf{S}_{1:i-1,*}$ randomly and calculate $\mathbf{F}_{\mathrm{BB}}[k]$ for each $k \in [1,K]$ according to \eqref{normalization}. 
            \STATE  Calculate $\hat{\textbf{H}}[k]=\textbf{H}[k]^H\textbf{H}[k]$ for each $k \in [1,K]$.
		
		\FOR{each $i \in [1,N_t]$}
		\STATE Design $\mathbf{S}_{i,*}$ to maximize \eqref{calculationS} by one-dimensional search. 
  		\ENDFOR
            \STATE  Calculate $\mathbf{F}_{\mathrm{RF}}$ according to \eqref{Frfdy}.
            \STATE  Calculate $\mathbf{F}_{\mathrm{BB}}[k]$ for each $k \in [1,K]$ according to \eqref{digitalprecoding} and \eqref{normalization}.
		\STATE Check convergence. If not, go back to step 3.
		\RETURN $\mathbf{F}_{\mathrm{RF}}$ and $\mathbf{F}_{\mathrm{BB}}[k]$ for each $k \in [1,K]$.
	\end{algorithmic}
\end{algorithm}

\section{Simulation Results And Analysis}
\label{section5}
In this section, we evaluate the performance of the proposed alternating optimization algorithms for different hybrid precoding structures in THz wideband communication systems. In the simulation, we consider a single user communication scenario that works at a center frequency of 300GHz. The bandwidth varies from 0.5GHz to 30GHz. As mentioned in Section \ref{section2}, ULA is deployed at the transmitter and the number of antennas is from 64 to 256. The number of RF chains is set to 16 at the transmitter and the number of data streams $N_s$ is 5 which satisfies $N_s<N^t_{RF}$. With regard to channel parameters, the number of clusters $N_{cl}$ is randomly generated within the range of 1 to 3, which fully considers the sparsity of the high-frequency channels. The number of rays $N_{ray}$ within each cluster is randomly set between 3 to 5. Moreover, the azimuth AODs/AoAs are uniformly distributed over $[-\frac{\pi}{3},\frac{\pi}{3}]$ with angular spread of $10^{\circ}$. The time delay $\tau_{m,n}$ is also assumed to follow uniform distribution which is described as $\tau_{m,n}\sim\mathcal{U}(0,1ns)$. The noise power $\sigma^2_n$ is set as 1. The simulation parameters are summarized in Table~\ref{parameters}.
\begin{table}[!htb]
	\caption{Simulation Parameters}
	\setlength{\tabcolsep}{0.7mm} 
	\renewcommand\arraystretch{1.2} 
	\centering
	\begin{tabular}{m{3.5cm}<{\centering}m{2.5cm}<{\centering}}
		\toprule
		\textbf{Parameter} & \textbf{Value}    \\ \midrule
		Center frequency $f_c$ & 300GHz           \\
            Bandwidth $B$ & 0.5GHz$\sim$30GHz       \\
		Number of antennas $N_t$ & 64$\sim$256 \\
            Number of RF chains $N^t_{RF}$ & 16 \\
            Number of data streams $N_s$ & 5 \\
            Number of clusters $N_{cl}$ & 1$\sim$3 \\
            Number of rays $N_{ray}$ & 3$\sim$5 \\
            AoDs/AoAs & $[-\frac{\pi}{3},\frac{\pi}{3}]$ \\
            Angle spread & $10^\circ$ \\
            Time delay $\tau_{m,n}$ & $\mathcal{U}(0,1ns)$ \\ \bottomrule
	\end{tabular}
	\label{parameters}
\end{table}

\subsection{Hybrid precoding Analysis in Fully Connected Structures}
\label{performanceF}
In this part, we evaluate the performance of our proposed alternating optimization algorithm in FC Structures(AlterOpt-FC). Several schemes are selected as comparisons: \textit{i}) Fully digital (FD) precoding which provides a performance upper bound; \textit{ii}) hybrid precoding with infinite PSs by SVD (IR-SVD) \cite{9262080}; \textit{iii}) hybrid precoding with low-resolution PSs (HPAC) \cite{sohrabi2017hybrid}; \textit{iv}) hybrid precoding with fixed PSs (FPHP) \cite{yu2018hardware}. IR-SVD in \cite{9262080} provides a low-complexity hybrid precoding with infinite PSs purely based on the array vectors. HPAC in \cite{sohrabi2017hybrid} proposes a heuristic hybrid percoding design to maximize the upper bound of specific efficiency. In \cite{yu2018hardware}, the authors design a new hybrid precoding structure which adopts fixed PSs and effectively improved spectral efficiency. In the following part, we will compare the performance of different schemes in terms of SNR, bandwidth, number of antennas, and PS resolution.

\begin{figure}[htbp]
	\centerline{\includegraphics[width=1.0\linewidth]{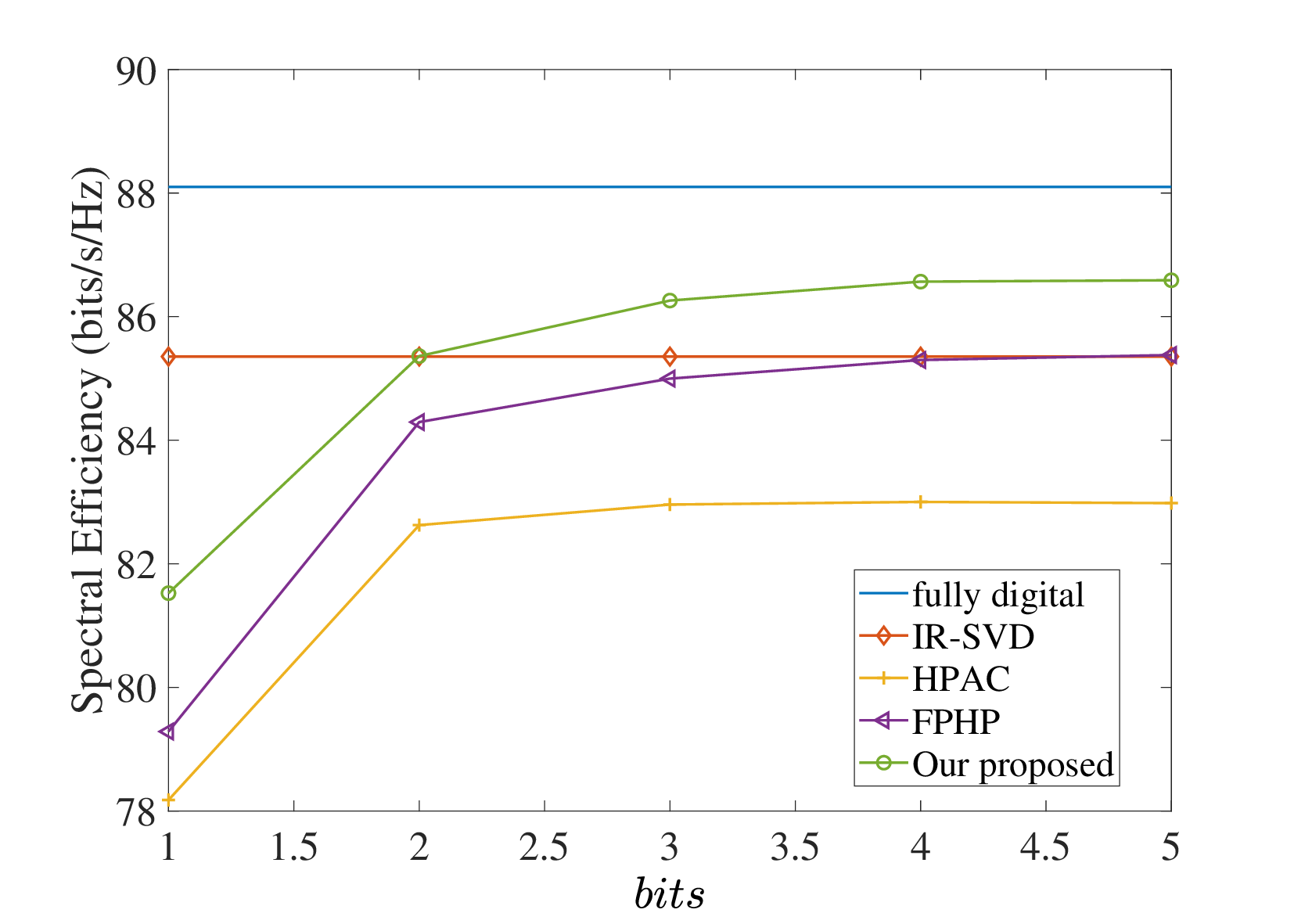}}
	\captionsetup{font=small}
	\caption{Spectral efficiency versus PS resolution, where $N_t=128$, $N_s=5$, $N^t_{RF}=16$, and $B=30\mathrm{GHz}$.}
	\label{bits}
\end{figure}

Fig.\ref{bits} illustrates the performance of our scheme and the comparisons versus PS resolution, where $N_t=128$, $N_s=5$,  $N^t_{RF}=16$, $SNR=20$, and $B=30\mathrm{GHz}$. We can see that the spectral efficiency of HPAC, FPHP, and our scheme is influenced heavily when the resolution is less than 3. With resolutions higher than 3 bits, there is a gradual improvement in performance as the resolution increases. This implies that it is practically feasible to use low-resolution PSs in real THz wideband MIMO systems. Moreover, the performance of our scheme is always better than the comparisons considering finite PSs, which demonstrates the effectiveness of our scheme at various resolutions. 

\begin{figure}[htbp]
	\centerline{\includegraphics[width=1.0\linewidth]{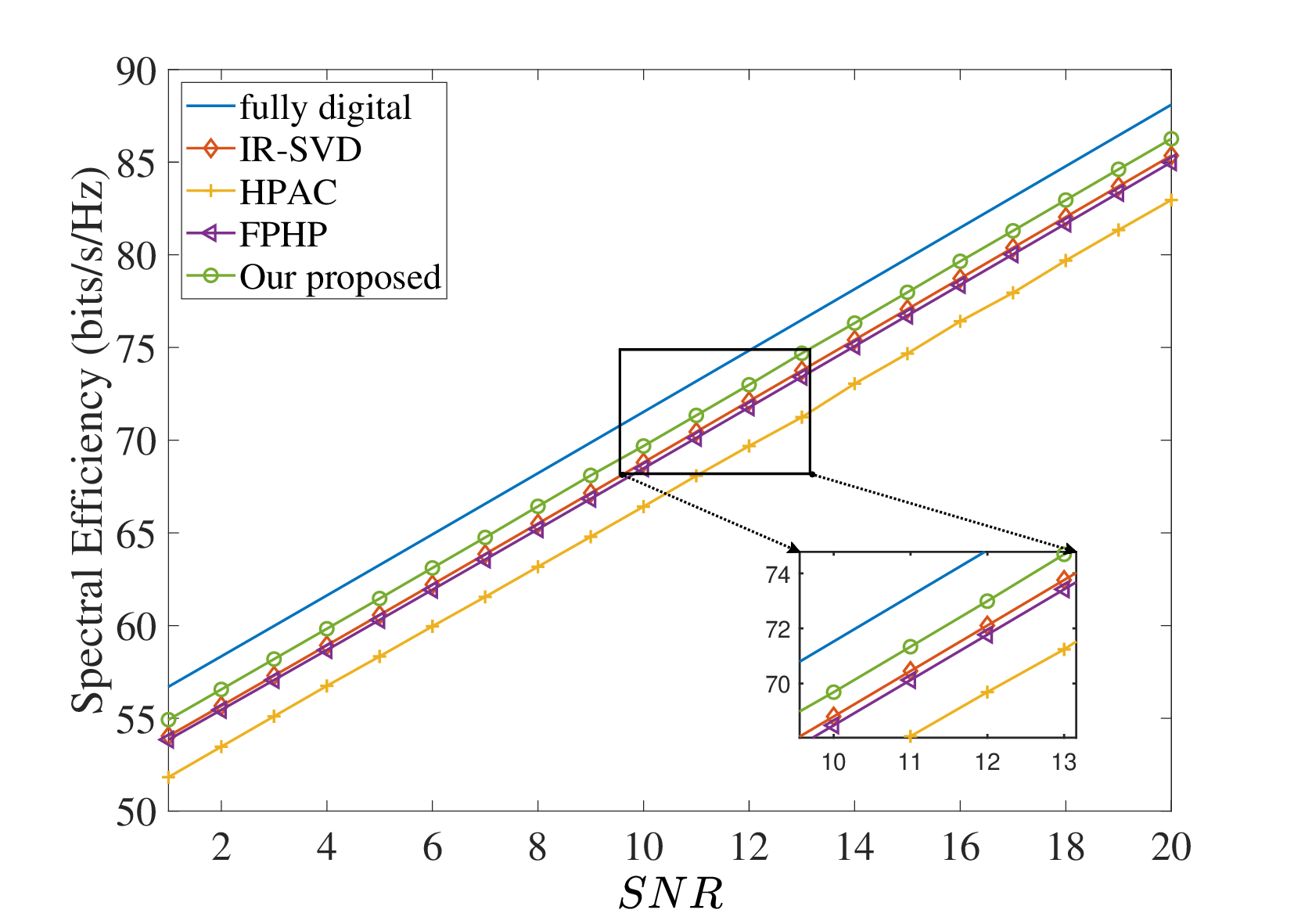}}
	\captionsetup{font=small}
	\caption{Spectral efficiency versus SNR, where $N_t=128$, $N_s=5$,  $N^t_{RF}=16$, and $B=30\mathrm{GHz}$.}
	\label{snr}
\end{figure}
Fig.\ref{snr} shows the average spectral efficiency versus SNR, where $N_t=128$, $N_s=5$,  $N^t_{RF}=16$, 3 bits resolution and $B=30\mathrm{GHz}$. It's obvious that our proposed algorithm always achieves better performance comparing IR-SVD, HPAC and FPHP under different SNR, approaching the performance of FD. Though IR-SVD adopts infinite resolution PSs, its performance is still worse than our proposed scheme. That's because the finite resolution schemes approach the performance of infinite resolution. Moreover, the performance of our proposed heuristic algorithm optimized for each PS is superior to that of the low-complexity SVD method.

\begin{figure}[htbp]
	\centerline{\includegraphics[width=1.0\linewidth]{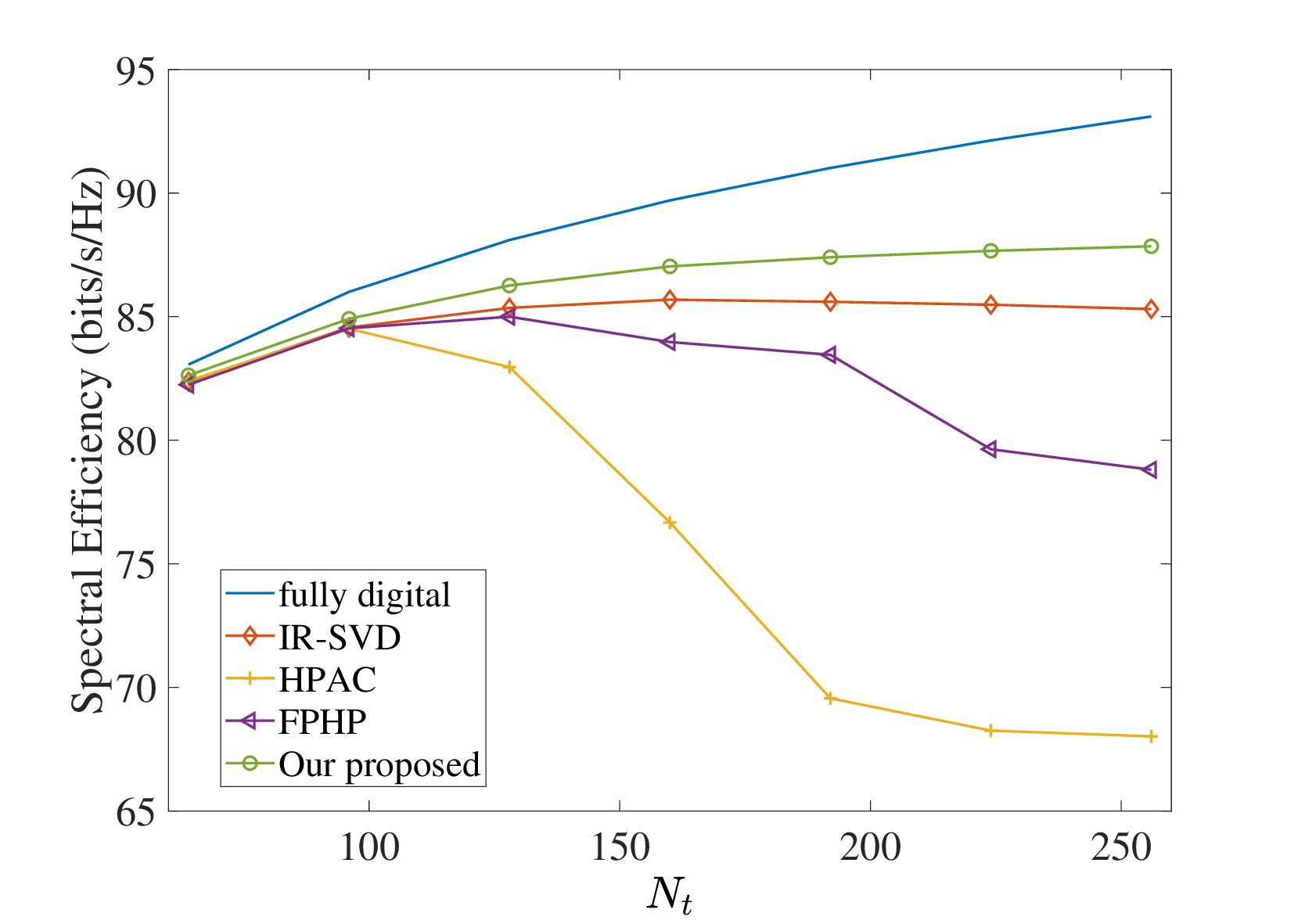}}
	\captionsetup{font=small}
	\caption{Spectral efficiency versus $N_t$, where $N_t=128$, $N_s=5$, $N^t_{RF}=16$, and $B=30\mathrm{GHz}$.}
	\label{Nt}
\end{figure}
In Fig.\ref{Nt}, we show the average spectral efficiency across varying numbers of $N_t$. An interesting phenomenon is that the performance of some schemes (HPAC and FPHP) decreases as the number of antennas increases. We believe this is due to the increase in $N_t$ causing the beams to become narrower, which leads to more significant beam squint. Meanwhile, our proposed scheme continues to show an upward trend and exhibits more pronounced performance advantages with large-scale antennas.

\begin{figure}[htbp]
	\centerline{\includegraphics[width=1.0\linewidth]{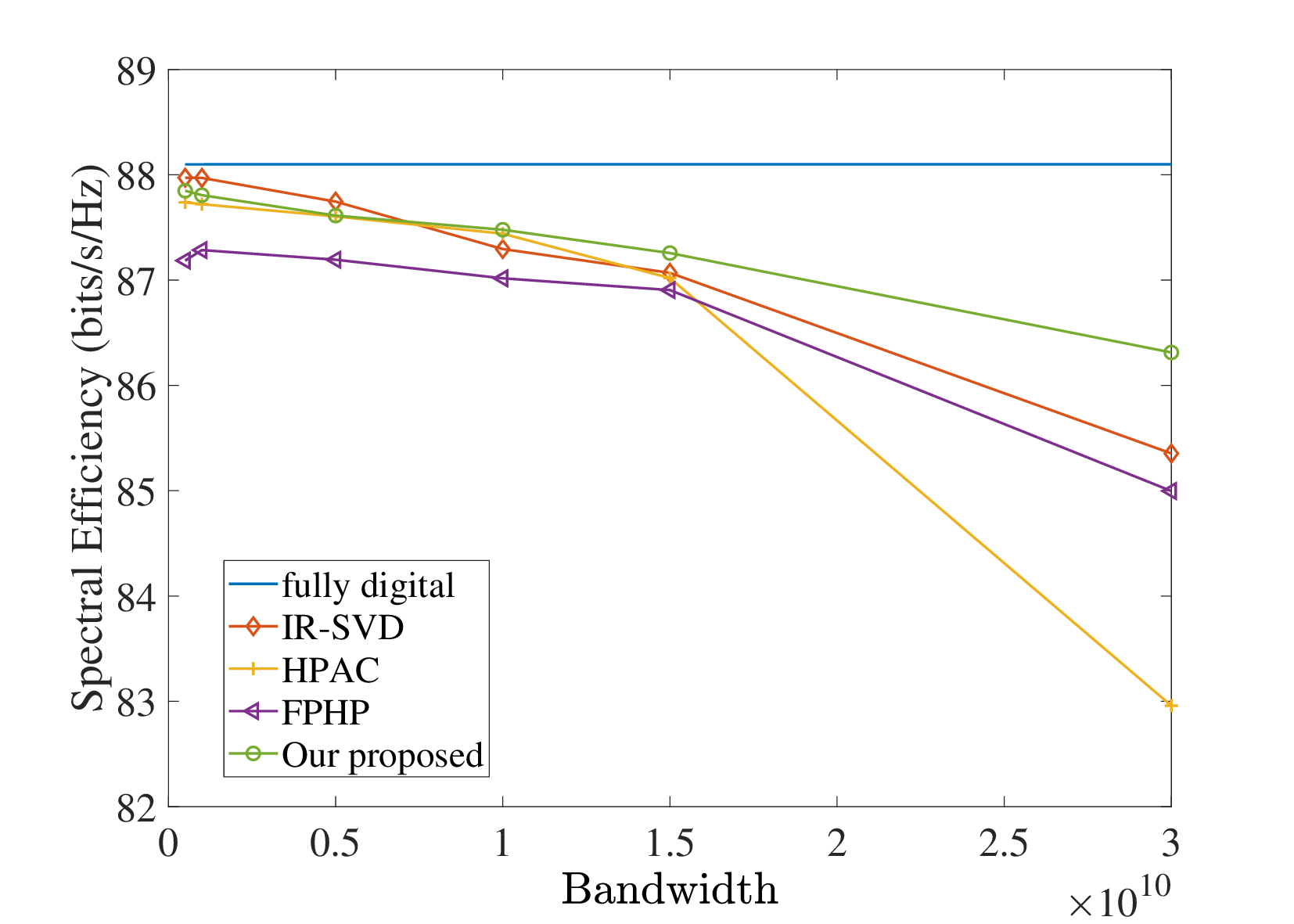}}
	\captionsetup{font=small}
	\caption{Spectral efficiency versus bandwidth, where $N_t=128$, $N_s=5$, $N^t_{RF}=16$, and $B=30\mathrm{GHz}$.}
	\label{bandwidth}
\end{figure}
Figure. \ref{bandwidth} illustrates the average specific efficiency of different schemes versus bandwidth. Aside from FD, the performance of other schemes declines as the bandwidth increases. Compared with other schemes, the decline in spectral efficiency of our scheme is more gradual, resulting in better performance under wideband.

\subsection{Hybrid precoding Analysis in Partially Connected Structures}
\label{performanceD}
We analyze the performance of our proposed alternating optimization algorithms in PC structures in this part. In this paper, we consider two kinds of PC structures : PC structures with fixed array and PC structures with dynamic array. In section \ref{four}, we propose two alternating optimization algorithms for the two structures: the alternating optimization algorithm for fixed subarray(Alteropt-PC) and the alternating optimization algorithm for dynamic subarray(DS-WB). For comparison, several schemes are selected: \textit{i}) Fully digital precoding (FD), which is the upper bound of all schemes; \textit{ii}) hybrid precoding for dynamic subarray with fixed phase shifter(DS-FPS)\cite{yan2022dynamic}; \textit{iii}) hybrid precoding with the group-connected mapping strategy\cite{yu2018hardware}. Next, we will illustrate the performance of our proposed algorithms from SNR and bandwidth.

\begin{figure}[htbp]
	\centerline{\includegraphics[width=1.0\linewidth]{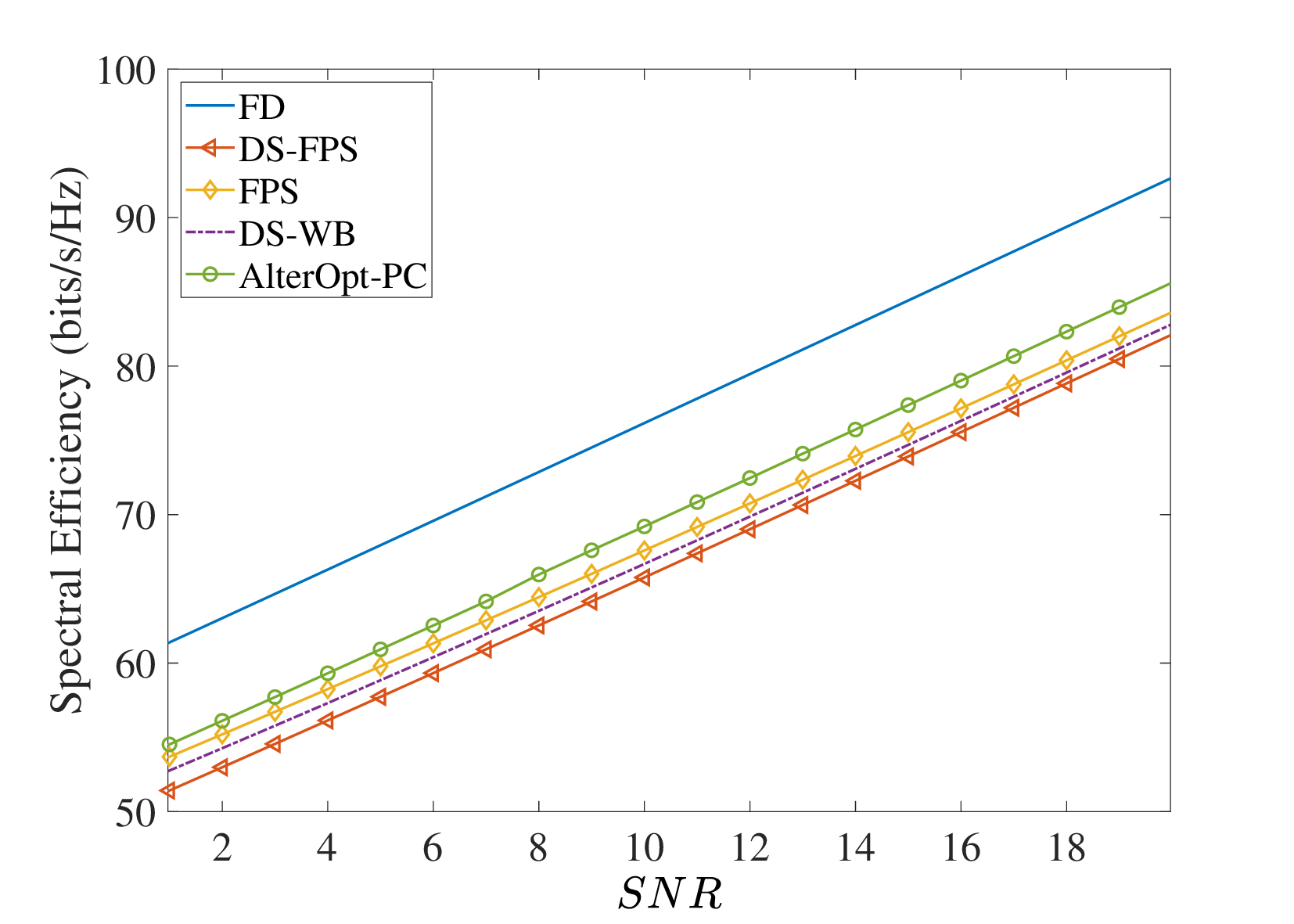}}
	\captionsetup{font=small}
	\caption{Spectral efficiency versus bandwidth, where $N_t=128$, $N_s=5$, $N^t_{RF}=16$, and $B=30\mathrm{GHz}$.}
	\label{fcsnr}
\end{figure}
Fig.\ref{fcsnr} shows the spectral efficiency versus SNR. It's obvious that AlterOpt-PC achieves the best performance compared with the comparisons under high bandwidth, which is closest to FD. Although PC with dynamic subarrays are more flexible, the advantages of DS-WB and DS-FPS are not obvious. One of the possible reasons is that our proposed AlterOpt-PC approaches global optima but the dynamic subarrays fall into local optima due to excessive degrees of freedom. But our proposed DS-WB achieves better performance versus different SNR compared with DS-FPS, which demonstrates our proposed alternating optimization algorithm achieves superior performance in PC with dynamic subarrays. Notably, FPS performance is superior to both DS-WB and DS-FPS. This is because FPS is not a strictly PC structure, as FPS allows each antenna to be connected to multiple fixed PSs. 

\begin{figure}[htbp]
	\centerline{\includegraphics[width=1.0\linewidth]{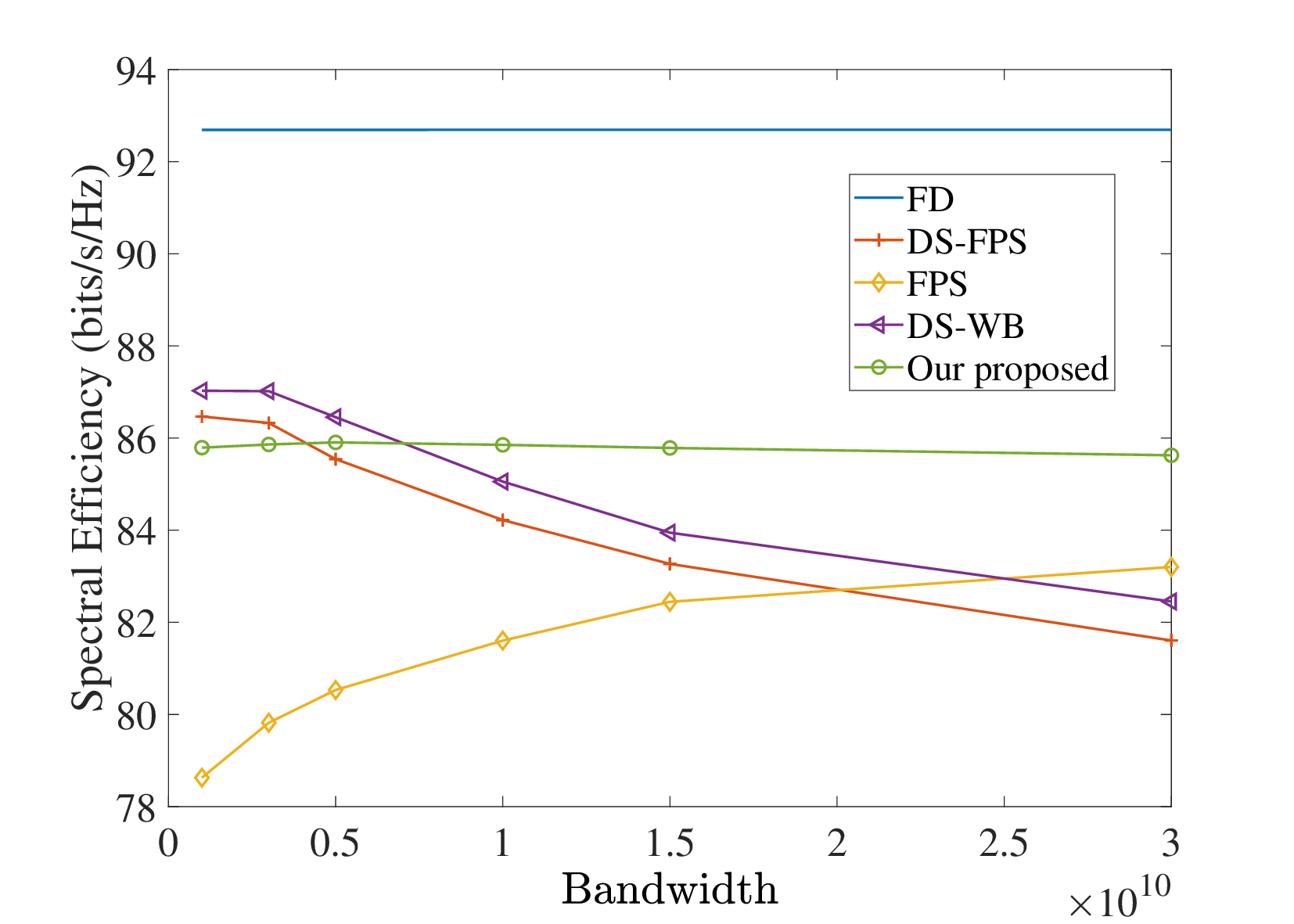}}
	\captionsetup{font=small}
	\caption{Spectral efficiency versus bandwidth, where $N_t=128$, $N_s=5$, $N^t_{RF}=16$, and $B=30\mathrm{GHz}$.}
	\label{fcBD}
\end{figure}
In Fig.\ref{fcBD}, we show the average spectral efficiency under different bandwidths to verify the performance of our proposed algorithms in both narrow band and wideband communication systems. We can observe that the performance curves of AlterOpt-PC and FD are stable, which means that the bandwidth has little impact on AlterOpt-PC, which makes AlterOpt-PC more suitable for wideband communications. DS-FPS and DS-WB achieve better performance compared with AlterOpt-PC in narrow band. With the bandwidth improving, the average spectral efficiency of DS-FPS and DS-WB decreases and becomes worse than AlterOpt-PC. But PC with fixed subarrays is a special case of PC with dynamic subarrays. That means the current schemes still have not yet reached its global optimum under wideband. Moreover, compared with DS-FPS, our proposed DS-WB achieves better performance versus different bandwidth, which demonstrates the effectiveness of DS-WB. FPS is another special scheme because the author in \cite{yu2018hardware} does not impose the 0 norm of $\|\bf{S}_\textit{i,*}\|_\mathsf{0} \leq \mathsf{1}$. So the total number of fixed PSs in FPS is not $N_t$. We can observe that the average spectral efficiency of FPS even rises with the bandwidth increasing. Because the number of fixed PSs increases, the performance approaches that of a FC structure.

\subsection{Convergence and Complexity Analysis}
\label{Complexity}
\begin{figure}[htbp]
        \centerline{\includegraphics[width=1.0\linewidth]{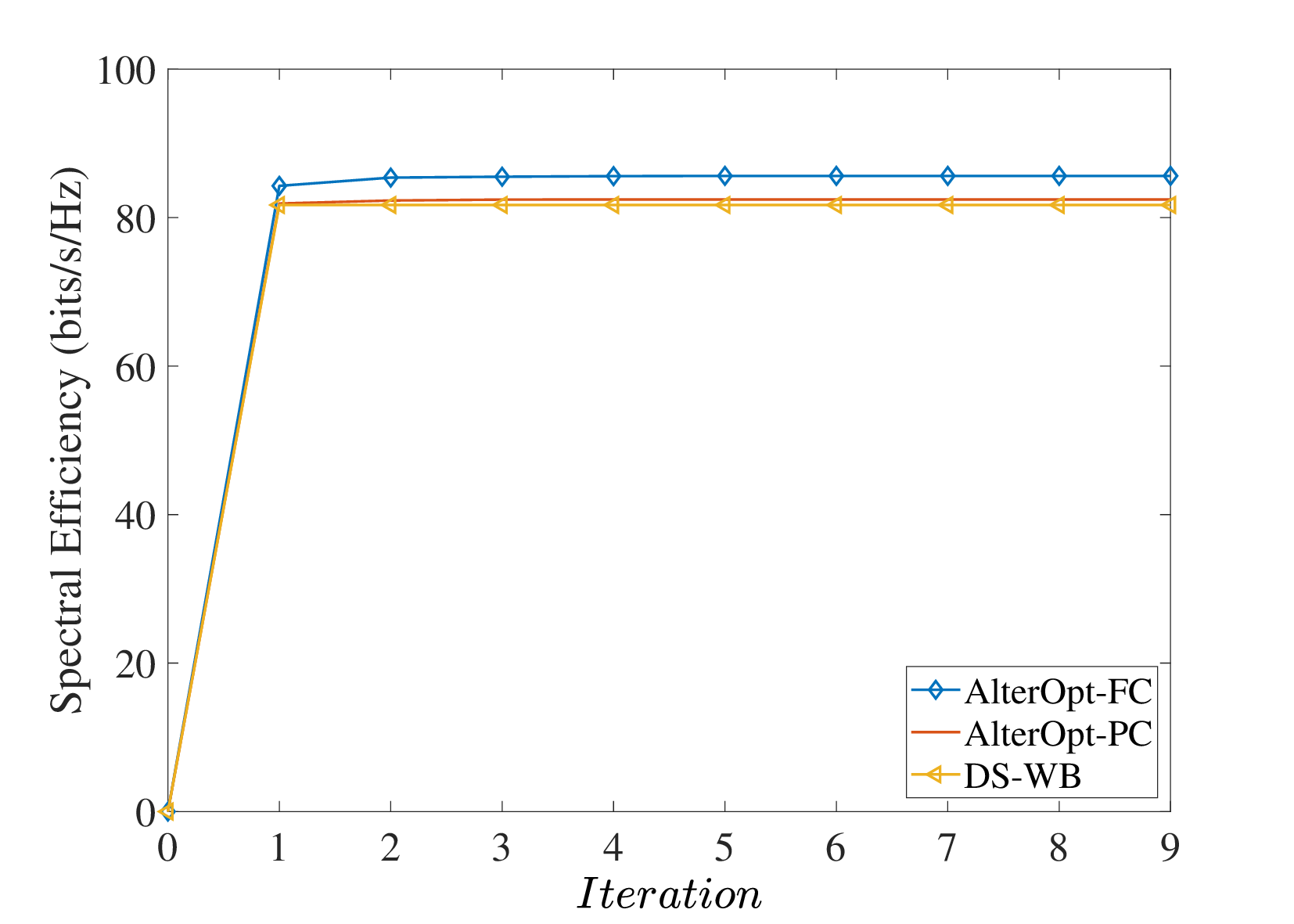}}
        \captionsetup[subfloat]{font=small,labelfont=rm}
	\caption{Convergence analysis in different structures.}
    \label{fig9}
\end{figure}
Fig.\ref{fig9} shows the convergence of our proposed AlterOpt-FC, AlterOpt-PC, and DS-WB. We evaluate the average spectral efficiency of the three
algorithms over all subcarriers versus the number of iterations. Fig.\ref{fig9} shows that our proposed AlterOpt-FC converges after two iteration. Numerical results indicate a slight increase with further iterations, but it is too subtle to appear in Fig.\ref{fig9}. Our proposed AlterOpt-PC and DS-WB for two kinds of PC structures even converge more quickly than AlterOpt-FC, requiring only one iteration shown in Fig.\ref{fig9}.

We also analyze the computational complexities of the three algorithms. Notably, there are relationships among system parameters: $N_s<N_{RF}^t<N_t$. So we have $\mathcal{O}(N_s)<\mathcal{O}(N_{RF}^t)\ll\mathcal{O}(N_t)$. For AlterOpt-FC, the main computational complexity is contributed by step 4, 6, and 8 in Algorithm 2. The step 4 requires a complexity of approximately $\mathcal{O}(KN_{RF}^t{N_t}^2)$, which is caused by the matrix multiplication. The step 6 requires a complexity of $\mathcal{O}(KN_{RF}^t{N_t}^2)$ which is produced by the phase optimization. The step 8 requires $\mathcal{O}(KN_{RF}^tN_tN_r)$ approximately due to SVD and matrix multiplication. The overall complexity of AlterOpt-FC is given by,
\begin{equation}
    \begin{aligned}	
    \label{ComplexityAF}
        \mathcal{O}(N_i(KN_{RF}^t{N_t}^2+KN_{RF}^t{N_t}^2+KN_{RF}^tN_tN_r))\\
        \approx\mathcal{O}(KN_iN_{RF}^t{N_t}^2).
    \end{aligned}  
\end{equation}
where $N_i$ represents the number of iteration. The main computational complexities of AlterOpt-PC are caused by steps 4, 6, and 10. The steps 4 and 6 all require $\mathcal{O}(K{N_t}^2)$. The step 10 requires $\mathcal{O}(KN_{RF}^tN_tN_r)$ operations approximately due to SVD and matrix multiplication. The overall complexity of AlterOpt-PC is given by,
\begin{equation}
    \begin{aligned}	
    \label{ComplexityAP}
        \mathcal{O}(N_i(K{N_t}^2+K{N_t}^2+KN_{RF}^tN_tN_r))\\\approx\mathcal{O}(KN_i{N_t}^2+KN_iN_{RF}^tN_tN_r).
    \end{aligned}  
\end{equation}
As for DS-WB, the main computational complexities are produced in steps 4, 6, and 7. The step 4 requires $\mathcal{O}(K{N_t}^2N^t_{RF}2^{b})$ operations. The operations in step 6 and 7 are $\mathcal{O}(KN_{RF}^tN_tN_r)$. The overall complexity of DS-WB is given by,
\begin{equation}
    \begin{aligned}	
    \label{ComplexityAP}
        &\mathcal{O}(N_i(K{N_t}^2N^t_{RF}2^{b}+KN_{RF}^tN_tN_r))\\
        &\approx\mathcal{O}(KN_i{N_t}^2N^t_{RF}2^{b}+KN_iN_{RF}^tN_tN_r).
    \end{aligned}  
\end{equation}

\section{Conclusion}
\label{five}
In this paper, we consider three different structures in massive MIMO and compare the advantages and disadvantages of each structure. For those structures, three alternating optimization algorithms are proposed to design the hybrid precoding with low-resolution PSs in the face of beam squint, which partially decouple the digital precoding matrix and the analog hybrid precoding matrix to improve the performance in wideband THz systems. A convergence and complexity analysis of the proposed algorithms is also provided. The simulation results demonstrate that the spectral efficiency of our proposed algorithm significantly outperforms the performance of the compared schemes with the same structure. Moreover, with the growth of bandwidth, our proposed algorithms become more pronounced.


{\appendix[Proof of the Optimal Phase]
\label{appendix1}
Based on \eqref{achievableratefully33}, we separate the contribution of the $n_{th}$ PS connected to the $m_{th}$ Rf chain and represent it as follows,
\begin{equation}
\begin{split}
    \varepsilon=&\frac{1}{K}\sum_{k=1}^K\log_2(\mathbf{D}_m[k]+\frac P{\sigma^2}\mathbf{F}_{RF,m,n}^H(\mathbf{G}_{m,n,\setminus n}[k]\mathbf{F}_{RF,m,\setminus n} \\
&-\mathbf{H}_{em,n}[k])+(\mathbf{F}_{RF,m,\setminus n}^H\mathbf{G}_{m,\setminus n,n}[k]-\mathbf{H}_{em,n}[k]^H)\times \\
&\mathbf{F}_{RF,m,n}-2real(\frac P{\sigma^2}\mathbf{H}_{em,\setminus n}[k]^H\mathbf{F}_{RF,m,\setminus n})+\frac P{\sigma^2}\times \\
&\mathbf{G}_{m,n,n}[k]+\frac P{\sigma^2}\mathbf{F}_{RF,m,\setminus n}^H\mathbf{G}_{m,\setminus n,\setminus n}[k]\mathbf{F}_{RF,m,\setminus n})
\end{split}
    \label{optimalphase}
\end{equation}
With the other PSs fixed, \eqref{optimalphase} is the complex function of $\mathbf{F}_{RF,m,n}$. However, we observe that $\mathbf{D}_m[k]$ and $\mathbf{G}_{m,n,n}[k]$ are both real numbers and the remaining terms appear in conjugate pairs in \eqref{optimalphase}. Let $\mathbf{F}_{RF,m,n}=x+yj$ and $\mathbf{F}_{RF,m,\setminus n}^H\mathbf{G}_{m,\setminus n,n}[k]-\mathbf{H}_{em,n}[k]^H=a+bj$, where $j$ is the imaginary unit. Then \eqref{optimalphase} can be further rewritten as,
\begin{equation}
\begin{split}
    \varepsilon=&\frac{1}{K}\sum_{k=1}^K\log_2(\mathbf{D}_m[k]+2\frac P{\sigma^2}(ax-by)-2real(\frac P{\sigma^2}\times\\ 
&\mathbf{H}_{em,\setminus n}[k]^H\mathbf{F}_{RF,m,\setminus n})+\frac P{\sigma^2}\mathbf{G}_{m,n,n}[k]+\frac P{\sigma^2}\times \\
&\mathbf{F}_{RF,m,\setminus n}^H\mathbf{G}_{m,\setminus n,\setminus n}[k]\mathbf{F}_{RF,m,\setminus n})
\end{split}
    \label{optimalphase1}
\end{equation}
Therefore, $\varepsilon$ can be considered a real function with its domain being $x^2+y^2=\frac{1}{N_t}$. We solve for the maximum points by calculating the first and second derivatives. We observe that the domain is a circle on a two-dimensional plane, and the extreme points of the function occur where the gradient vector is parallel to the normal of the circle. Thus, the following relationship exists,
\begin{equation}
    y\frac{\partial \varepsilon}{\partial x} = x\frac{\partial \varepsilon}{\partial y}
    \label{optimalphase2}
\end{equation}
The partial derivatives $\frac{\partial \varepsilon}{\partial x}$ and $\frac{\partial \varepsilon}{\partial y}$ are given by,
\begin{equation}
    \begin{split}
        \frac{\partial \varepsilon}{\partial x} = \frac{1}{K}\sum_{k=1}^K\frac{2\frac P{\sigma^2}a}{ln2\mathbf{M}[k]}\\   
        \frac{\partial \varepsilon}{\partial y} = \frac{1}{K}\sum_{k=1}^K-\frac{2\frac P{\sigma^2}b}{ln2\mathbf{M}[k]}
    \end{split}
    \label{optimalphase3}
\end{equation}
We consider that the contribution of a single PS to the overall spectral efficiency is minimal. Therefore, we assume that $\mathbf{M}[k]$ remains approximately unchanged before and after updating a particular PS. According to \eqref{optimalphase2}, \eqref{optimalphase3} and the constant modulus constraint, we can obtain \eqref{vartheta_1}.

\bibliographystyle{IEEEtran}
\bibliography{ref}

\begin{thebibliography}{10}
\providecommand{\url}[1]{#1}
\csname url@samestyle\endcsname
\providecommand{\newblock}{\relax}
\providecommand{\bibinfo}[2]{#2}
\providecommand{\BIBentrySTDinterwordspacing}{\spaceskip=0pt\relax}
\providecommand{\BIBentryALTinterwordstretchfactor}{4}
\providecommand{\BIBentryALTinterwordspacing}{\spaceskip=\fontdimen2\font plus
\BIBentryALTinterwordstretchfactor\fontdimen3\font minus
  \fontdimen4\font\relax}
\providecommand{\BIBforeignlanguage}[2]{{%
\expandafter\ifx\csname l@#1\endcsname\relax
\typeout{** WARNING: IEEEtran.bst: No hyphenation pattern has been}%
\typeout{** loaded for the language `#1'. Using the pattern for}%
\typeout{** the default language instead.}%
\else
\language=\csname l@#1\endcsname
\fi
#2}}
\providecommand{\BIBdecl}{\relax}
\BIBdecl

\bibitem{wang2023road}
C.-X. Wang, X.~You, X.~Gao, X.~Zhu, Z.~Li, C.~Zhang, H.~Wang, Y.~Huang,
  Y.~Chen, H.~Haas, J.~S. Thompson, E.~G. Larsson, M.~D. Renzo, W.~Tong,
  P.~Zhu, X.~Shen, H.~V. Poor, and L.~Hanzo, ``On the road to {6G}: Visions,
  requirements, key technologies, and testbeds,'' \emph{IEEE Commun. Surveys
  Tut.}, vol.~25, no.~2, pp. 905--974, 2023.

\bibitem{yang20196g}
P.~Yang, Y.~Xiao, M.~Xiao, and S.~Li, ``6g wireless communications: Vision and
  potential techniques,'' \emph{IEEE Netw.}, vol.~33, no.~4, pp. 70--75, 2019.

\bibitem{10360222}
Y.~Sun, C.~Yang, and M.~Peng, ``Subarray-based hybrid-field channel estimation
  for terahertz wideband {UM-MIMO} systems without prior location knowledge,''
  \emph{IEEE Trans. Veh. Technol.}, vol.~73, no.~5, pp. 7363--7367, 2024.

\bibitem{10045774}
B.~Ning, Z.~Tian, W.~Mei, Z.~Chen, C.~Han, S.~Li, J.~Yuan, and R.~Zhang,
  ``Beamforming technologies for ultra-massive mimo in terahertz
  communications,'' \emph{IEEE Open J. Commun. Soc.}, vol.~4, pp. 614--658,
  2023.

\bibitem{10105223}
Z.~Liu, C.~Yang, Y.~Sun, and M.~Peng, ``Closed-form model for performance
  analysis of {THz} joint radar-communication systems,'' \emph{IEEE Trans.
  Wirel. Commun.}, vol.~22, no.~12, pp. 8694--8706, 2023.

\bibitem{akyildiz2022terahertz}
I.~F. Akyildiz, C.~Han, Z.~Hu, S.~Nie, and J.~M. Jornet, ``Terahertz band
  communication: An old problem revisited and research directions for the next
  decade,'' \emph{IEEE Trans. Commun.}, vol.~70, no.~6, pp. 4250--4285, 2022.

\bibitem{9398864}
F.~Gao, B.~Wang, C.~Xing, J.~An, and G.~Y. Li, ``Wideband beamforming for
  hybrid massive mimo terahertz communications,'' \emph{IEEE J. Sel. Areas
  Commun.}, vol.~39, no.~6, pp. 1725--1740, 2021.

\bibitem{9216613}
A.~Faisal, H.~Sarieddeen, H.~Dahrouj, T.~Y. Al-Naffouri, and M.-S. Alouini,
  ``Ultramassive mimo systems at terahertz bands: Prospects and challenges,''
  \emph{IEEE Veh. Technol. Mag.}, vol.~15, no.~4, pp. 33--42, 2020.

\bibitem{9262080}
Y.~Chen, Y.~Xiong, D.~Chen, T.~Jiang, S.~X. Ng, and L.~Hanzo, ``Hybrid
  precoding for wideband millimeter wave mimo systems in the face of beam
  squint,'' \emph{IEEE Trans. Wireless Commun.}, vol.~20, no.~3, pp.
  1847--1860, 2021.

\bibitem{wang2023sensing}
Y.~Wang, C.~Yang, Z.~Ren, Y.~Sun, and M.~Peng, ``Sensing-aided hybrid precoding
  for efficient terahertz wideband communications in multiuser high-data-rate
  iot,'' \emph{IEEE Internet of Things J.}, vol.~11, no.~5, pp. 8253--8267,
  2024.

\bibitem{el2014spatially}
O.~El~Ayach, S.~Rajagopal, S.~Abu-Surra, Z.~Pi, and R.~W. Heath, ``Spatially
  sparse precoding in millimeter wave mimo systems,'' \emph{IEEE Trans.
  Wireless Commun.}, vol.~13, no.~3, pp. 1499--1513, 2014.

\bibitem{lee2014hybrid}
Y.-Y. Lee, C.-H. Wang, and Y.-H. Huang, ``A hybrid rf/baseband precoding
  processor based on parallel-index-selection matrix-inversion-bypass
  simultaneous orthogonal matching pursuit for millimeter wave mimo systems,''
  \emph{IEEE Trans. Signal Process.}, vol.~63, no.~2, pp. 305--317, 2014.

\bibitem{yu2016alternating}
X.~Yu, J.-C. Shen, J.~Zhang, and K.~B. Letaief, ``Alternating minimization
  algorithms for hybrid precoding in millimeter wave mimo systems,'' \emph{IEEE
  J. Sel. Topics Signal Process.}, vol.~10, no.~3, pp. 485--500, 2016.

\bibitem{li2017hybrid}
A.~Li and C.~Masouros, ``Hybrid precoding and combining design for
  millimeter-wave multi-user mimo based on {SVD},'' in \emph{Proc. IEEE Int.
  Conf. Commun. (ICC)}, 2017, pp. 1--6.

\bibitem{zhang2018svd}
D.~Zhang, P.~Pan, R.~You, and H.~Wang, ``{SVD}-based low-complexity hybrid
  precoding for millimeter-wave {MIMO} systems,'' \emph{IEEE Commun. Lett.},
  vol.~22, no.~10, pp. 2176--2179, 2018.

\bibitem{10438529}
L.~Piotto, G.~D. Filippi, M.~M. Pirbazari, and A.~Mazzanti, ``Compact d-band
  passive phase shifters with fine and coarse control steps in bicmos-55nm,''
  in \emph{2024 IEEE 24th Topical Meeting on Silicon Monolithic Integrated
  Circuits in RF Systems (SiRF)}, 2024, pp. 25--28.

\bibitem{10159567}
K.~Rasilainen, T.~D. Phan, M.~Berg, A.~Pärssinen, and P.~J. Soh, ``Hardware
  aspects of sub-thz antennas and reconfigurable intelligent surfaces for 6g
  communications,'' \emph{IEEE J. Sel. Areas Commun.}, vol.~41, no.~8, pp.
  2530--2546, 2023.

\bibitem{he2023energy}
Y.~He, M.~Shen, R.~Wang, and X.~Liu, ``Energy efficient hybrid precoder for
  cell-free wideband mmwave massive {MIMO} systems,'' \emph{IEEE Commun.
  Lett.}, vol.~27, no.~9, pp. 2491--2495, 2023.

\bibitem{zhu2023max}
W.~Zhu, H.~Tuan, E.~Dutkiewicz, H.~Poor, and L.~Hanzo, ``Max-min rate
  optimization of low-complexity hybrid multi-user beamforming maintaining
  rate-fairness,'' \emph{IEEE Trans. Wireless Commun.}, vol.~23, no.~6, pp.
  5648--5662, 2024.

\bibitem{nie2023spectrum}
W.~Nie, M.~Liu, J.~Chen, W.~Tan, and C.~Li, ``Spectrum and energy efficiency of
  massive {MIMO} for hybrid architectures with phase shifter and switches in
  iot networks,'' \emph{IEEE Internet Things J.}, vol.~11, no.~6, pp.
  9656--9667, 2024.

\bibitem{sohrabi2016hybrid}
F.~Sohrabi and W.~Yu, ``Hybrid digital and analog beamforming design for
  large-scale antenna arrays,'' \emph{IEEE J. Sel. Topics Signal Process.},
  vol.~10, no.~3, pp. 501--513, 2016.

\bibitem{chen2017hybrid}
J.-C. Chen, ``Hybrid beamforming with discrete phase shifters for
  millimeter-wave massive mimo systems,'' \emph{IEEE Trans. Veh. Technol.},
  vol.~66, no.~8, pp. 7604--7608, 2017.

\bibitem{lyu2021lattice}
S.~Lyu, Z.~Wang, Z.~Gao, H.~He, and L.~Hanzo, ``Lattice-based mmwave hybrid
  beamforming,'' \emph{IEEE Trans. Commun.}, vol.~69, no.~7, pp. 4907--4920,
  2021.

\bibitem{sohrabi2017hybrid}
F.~Sohrabi and W.~Yu, ``Hybrid analog and digital beamforming for mmwave ofdm
  large-scale antenna arrays,'' \emph{IEEE J. Sel. Areas Commun.}, vol.~35,
  no.~7, pp. 1432--1443, 2017.

\bibitem{zilli2021constrained}
G.~M. Zilli and W.-P. Zhu, ``Constrained tensor decomposition-based hybrid
  beamforming for mmwave massive mimo-ofdm communication systems,'' \emph{IEEE
  Trans. Veh. Technol.}, vol.~70, no.~6, pp. 5775--5788, 2021.

\bibitem{li2020dynamic}
H.~Li, M.~Li, Q.~Liu, and A.~L. Swindlehurst, ``Dynamic hybrid beamforming with
  low-resolution pss for wideband mmwave mimo-ofdm systems,'' \emph{IEEE J.
  Sel. Areas Commun.}, vol.~38, no.~9, pp. 2168--2181, 2020.

\bibitem{yan2022dynamic}
L.~Yan, C.~Han, N.~Yang, and J.~Yuan, ``Dynamic-subarray with fixed phase
  shifters for energy-efficient terahertz hybrid beamforming under partial
  csi,'' \emph{IEEE Trans. Wireless Commun.}, vol.~22, no.~5, pp. 3231--3245,
  2023.

\bibitem{yu2018hardware}
X.~Yu, J.~Zhang, and K.~B. Letaief, ``A hardware-efficient analog network
  structure for hybrid precoding in millimeter wave systems,'' \emph{IEEE J.
  Sel. Topics Signal Process.}, vol.~12, no.~2, pp. 282--297, 2018.

\bibitem{10333083}
Y.~Gao, H.~Hu, B.~Chen, J.~Tian, and S.~Lei, ``Hybrid precoding for mitigating
  the beam squint in wideband mmwave mimo system,'' \emph{IEEE Wireless Commun.
  Lett.}, vol.~13, no.~3, pp. 602--606, 2024.

\bibitem{zhang2018hybrid}
J.~Zhang, Y.~Huang, J.~Wang, and L.~Yang, ``Hybrid precoding for wideband
  millimeter-wave systems with finite resolution phase shifters,'' \emph{IEEE
  Trans. Veh. Technol.}, vol.~67, no.~11, pp. 11\,285--11\,290, 2018.

\bibitem{9908559}
Q.~Yuan, H.~Liu, M.~Xu, Y.~Wu, L.~Xiao, and T.~Jiang, ``Deep learning-based
  hybrid precoding for terahertz massive mimo communication with beam squint,''
  \emph{IEEE Commun. Lett.}, vol.~27, no.~1, pp. 175--179, 2023.

\bibitem{linx2017hybrid}
N.~Li, Z.~Wei, H.~Yang, X.~Zhang, and D.~Yang, ``Hybrid precoding for mmwave
  massive mimo systems with partially connected structure,'' \emph{IEEE
  Access}, vol.~5, pp. 15\,142--15\,151, 2017.

\bibitem{park2017dynamic}
S.~Park, A.~Alkhateeb, and R.~W. Heath, ``Dynamic subarrays for hybrid
  precoding in wideband mmwave mimo systems,'' \emph{IEEE Trans. Wireless
  Commun.}, vol.~16, no.~5, pp. 2907--2920, 2017.

\bibitem{7394147}
A.~Alkhateeb, Y.-H. Nam, J.~Zhang, and R.~W. Heath, ``Massive mimo combining
  with switches,'' \emph{IEEE Wireless Commun. Lett.}, vol.~5, no.~3, pp.
  232--235, 2016.

\bibitem{9433528}
J.~Wang, X.~Zhang, X.~Shi, and J.~Song, ``Higher spectral efficiency for mmwave
  mimo: Enabling techniques and precoder designs,'' \emph{IEEE Commun. Mag.},
  vol.~59, no.~4, pp. 116--122, 2021.

\end{thebibliography}



\newpage

\section{Biography Section}
 
\vspace{11pt}


\vspace{11pt}


\vfill
\end{document}